\documentclass[journal]{IEEEtran}
\usepackage{tikz}
\usetikzlibrary{calc}

\ifCLASSINFOpdf
\else
\fi
\newcommand{\bvec}[1]{\ensuremath{{\boldsymbol{#1}}}}
\newcommand{\dvec}[1]{\ensuremath{{\bvec{\vec{#1}}}}}
\newcommand{\pvec}[1]{\vec{#1}\mkern2mu\vphantom{#1}'}

\usepackage{mathtools}
\DeclarePairedDelimiterX{\infdivx}[2]{(}{)}{%
  #1\;\delimsize\|\;#2%
}
\newcommand{\DKL}{D_\text{KL}\infdivx}
\newcommand{\Dll}[2]{\ensuremath{D(\boldsymbol{\lambda}_{#1} | \boldsymbol{\lambda}_{#2})}}
\newcommand{\Dnl}[2]{\ensuremath{D(\boldsymbol{n}_{#1} | \boldsymbol{\lambda}_{#2})}}
\newcommand{\Lshape}{\ensuremath{\mathsf{L}}-shape}
\usepackage{xcolor}

\newcommand{\Nfourt}{\ensuremath{N_{64}(t)}}
\newcommand{\Nthreet}{\ensuremath{N_{63}(t)}}
\newcommand{\Nfour}{\ensuremath{N_{64}}}
\newcommand{\Nthree}{\ensuremath{N_{63}}}

\newcommand{\Afourt}{\ensuremath{A_{64}(t)}}
\newcommand{\Afour}{\ensuremath{A_{64}}}

\newcommand{\dd}{\ensuremath{\text{d}}}
\newcommand{\Dtcook}{\ensuremath{\Delta t_\text{cook}}}
\newcommand{\Dtcool}{\ensuremath{\Delta t_\text{cool}}}

\usepackage{enumitem}

\usepackage{amsmath}
\usepackage{amssymb}
\usepackage{url}
\usepackage{hyperref}

\begin{document}
\IEEEpubid{\begin{minipage}{0.85\textwidth}\ \\[12pt]\\ \\ \\ \centering
  This work has been submitted to the IEEE for possible publication. Copyright may be transferred without notice, after which this version may no longer be accessible.
\end{minipage}}
%
\title{Design and deployment of radiological point-source arrays for the emulation of continuous distributed sources}
%
%
%
\author{
    Jayson~R.~Vavrek,
    C.~Corey~Hines,
    Mark~S.~Bandstra,
    Daniel~Hellfeld,
    Maddison~A.~Heine,
    Zachariah~M.~Heiden,
    Nick~R.~Mann,
    Brian~J.~Quiter,
    and Tenzing~H.Y.~Joshi
    \thanks{
        JRV, MSB, DH, BJQ, and THYJ are with the Applied Nuclear Physics program at Lawrence Berkeley National Laboratory.
        CCH, MAH, and ZMH are with the Nuclear Science Center at Washington State University.
        NRM is with the National and Homeland Security Research Program at Idaho National Laboratory.
    }%
}

\markboth{IEEE Transactions on Nuclear Science}%
{Vavrek \MakeLowercase{\textit{et al.}}}
%



\maketitle

\begin{abstract}
We demonstrate a method for using arrays of point sources that emulate---when measured from a standoff of at least several meters---distributed gamma-ray sources, and present results using this method from outdoor aerial measurements of several planar arrays each comprising up to $100$ ${\sim}7$~mCi Cu-64 sealed sources.
The method relies on the Poisson deviance to statistically test whether the array source ``looks like'' its continuous analogue to a particular gamma-ray detector given the counts recorded as the detector moves about 3D space.
We use this deviance metric to design eight different mock distributed sources, ranging in complexity from a $36\times36$~m uniform square grid of sources to a configuration where regions of higher and zero activity are superimposed on a uniform baseline.
We then detail the design, manufacture, and testing of the ${\sim} 7$~mCi Cu-64 sealed sources at the Washington State University research reactor, and their deployment during the aerial measurement campaign.
We show the results of two such measurements, in which approximate source shapes and qualitative source intensities can be seen.
Operationally, we find that the point-source array technique provides high source placement accuracy and ease of quantifying the true source configuration, scalability to source dimensions of ${\lesssim}100$~m, ease of reconfiguration and removal, and relatively low dose to personnel.
Finally, we consider potential improvements and generalizations of the point-source array technique for future measurement campaigns.
\end{abstract}


%
\IEEEpeerreviewmaketitle

\section{Introduction}
%
%
%
%
\IEEEPARstart{Q}{uantitatively} mapping continuous distributed radiological sources is important for radiological emergency response, whether the cause of the radioactive release is accidental (e.g., contamination from a reactor accident), or intentional (e.g., nuclear warfare).
Testing and validating mapping and imaging algorithms for such distributed sources is challenging, however, for three related reasons.
First, it is difficult to manufacture and deploy truly distributed radiation sources---radioactive material would have to be powdered, aerosolized, or dissolved, which can present a substantial human and environmental safety hazard~\cite{beckman2020robotic}, especially if the radioactive material were ingested or inhaled.
For instance, measurements of dispersed activated KBr~\cite{pennington2020microscale} or La$_2$O$_3$~\cite{green2016overview} powder in ${\sim}1$~Ci ($37$~GBq) quantities require substantial personal protective equipment (PPE)~\cite[Annex~3]{dhs2017guidance} and large standoff distances due to high concentrations of airborne and deposited radioactivity.
Second, powdered, aerosolized, or dissolved radionuclides cannot be easily and safely reconfigured, making it difficult to rapidly test multiple source configurations, to transfer the source material back to a laboratory for later assay, or---if the source half-life is long---to ensure that all radioactive material is removed from the environment.
Third, producing truly continuous distributions of radioactive material with known ground truth patterns is also difficult---the deposition of powdered, aerosolized, or dissolved materials may deviate from the intended pattern due to factors such as changing winds or uneven mixing due to the mechanical variability of the depositor in inclement weather~\cite{beckman2020robotic}.
Post-deposition ground truth activity assays are possible using collimated high-purity germanium (HPGe)~\cite{swinney2018methodology} or cerium bromide (CeBr$_3$)~\cite{simerl2021contamination} detectors, but these measurements require very close proximity to the source (increasing dose and often disturbing contaminated soil) and are limited to small (${\lesssim}1$~m$^2$) areas in a single measurement, and thus are difficult to use for rapidly mapping large distributed sources spanning hundreds or thousands of square meters.
While remotely- or autonomously-controlled ground robots could be used to carry the detectors used for the ground truth measurements and this would mitigate dose concerns, their use would create additional complications such as ensuring the robots did not disturb the source distributions or become radiologically contaminated themselves.

Instead, in this work, we present a method for emulating truly continuous distributed radiological sources with arrays of sealed point sources, which are easily ground-truthable, re-configurable, and removable.
The method is founded on the Poisson deviance, and uses this metric to ask how much the array source ``looks like'' its continuous source analogue, for a given detector and trajectory.
In presenting these calculations, we will rely on the concept of ``spoofing'', wherein a successful spoof is one where the fake array of sources cannot be distinguished from the continuous source analogue.  
Using this metric as a guide, we simulate unmanned aerial system (UAS) borne gamma-ray measurements (using the NG-LAMP~\cite{Pavlovsky2019} and MiniPRISM~\cite{pavlovsky2019miniprism} detectors) of eight different planar array source patterns of ${\sim}500$~mCi of Cu-64 each.
Finally, we detail the design and deployment of $300$ ${\sim}7$~mCi ($259$~MBq) Cu-64 sources in these eight array source patterns during an August 2021 outdoor distributed sources measurement campaign at Washington State University (WSU).
These measurements will form the basis of an upcoming study comparing quantitative MAP-EM~\cite{shepp1982maximum, hellfeld2019gamma} reconstructions of the source distributions against known ground truth.

\section{Methods}
\subsection{Mathematical framework}

The degree to which a detector can distinguish a continuous source from an array depends on a number of factors, including the detector trajectory, the detector response, and the source itself.
We consider a detector trajectory $\dvec{r} \in \mathbb{R}^{I \times 3}$ with dwell times $\bvec{t} \in \mathbb{R}_{+}^{I}$, where~$I$ is the number of individual time-binned measurements made over the course of the full measurement.\footnote{We use boldface italic to denote arbitrary vectors $\bvec{v} \in \mathbb{R}^N$, vector arrow notation for a single 3D vector $\vec{v} \in \mathbb{R}^3$, and both to denote a collection of~$N$ vectors in 3D, $\dvec{v} \in \mathbb{R}^{N \times 3}$.}\textsuperscript{,}\footnote{For simplicity we present the mathematical framework for a single detector, but it can be readily generalized to $J$ detector elements.}
In addition to its position vector, the detector also has a vector of orientations~$\textbf{a}$, each element of which can be described by a quaternion or rotation matrix.
Each pair of position and orientation, known as a ``pose'', in turn
influences the detector response or ``effective area'' $\bvec{\eta}(\textbf{a}, \dvec{r} - \pvec{r})$ to a point source of radiation located at~$\pvec{r}$.

A truly continuous distributed source of radiation has a per-volume intensity distribution $w(\pvec{r}),\, \forall\, \pvec{r} \in \mathbb{R}^3$.
Neglecting attenuation from any intervening material, the expected number of photopeak counts detected at each pose $\bvec{\lambda} \in \mathbb{R}_{+}^{I}$ involves integrating over the source distribution:
\begin{align}\label{eq:lambda_cts}
    \bvec{\lambda} = \int_{\mathbb{R}^3} \frac{w(\pvec{r}) \bvec{\eta}(\textbf{a}, \dvec{r} - \pvec{r}) \bvec{t}}{4 \pi |\dvec{r} - \pvec{r}|^2} \text{d}^3\pvec{r}.
\end{align}
Given that this work focuses on planar sources, we note that for an isotropic detector at a constant height~$z=h$ above the center of a uniform circular plane source of radius $R$ at $z=0$ with an activity density $w(\pvec{r}) = \delta(z) w_0$, $0 \leq |\pvec{r}| \leq R$, we have the analytical solution
\begin{align}\label{eq:lambda_plane}
    \lambda_i = \frac{w_0 \eta_i t_i}{4} \log\left( 1 + R^2/h^2 \right),\quad i=1, 2, \ldots, I.
\end{align}
We note that Eq.~\ref{eq:lambda_plane} decreases more slowly with height $h$ compared to the familiar $1/h^2$ behavior that occurs for point sources as well as when $h \gg R$.

For arbitrary distributions $w(\pvec{r})$, however, the integral in Eq.~\ref{eq:lambda_cts} may be difficult or impossible to evaluate analytically~(see, e.g., Refs.~\cite{hubbell1960radiation}, \cite[Appendix~A]{proctor1997aerial}, \cite[Chapter~4]{auxier1959experimental}).
A more computationally-oriented approach suitable for arbitrary distributions involves discretizing the distribution $w(\pvec{r})$ into $K$ point sources, in which case the expected number of counts is
\begin{align}\label{eq:lambda_arr}
    \bvec{\lambda} = \sum_{k=1}^{K} \frac{w_k \bvec{\eta}(\textbf{a}, \dvec{r} - \pvec{r}_k) \bvec{t}}{4\pi |\dvec{r} - \pvec{r}_k|^2}.
\end{align}
Here the $w_k$ would typically be chosen by voxelizing (a bounded subset of) $\mathbb{R}^3$, evaluating $w(\pvec{r})$ at each of the $K$ voxel centers, and multiplying by each voxel volume.
As $K$ increases and the voxel size decreases (for a fixed total volume), the fidelity to Eq.~\ref{eq:lambda_cts} increases, but so too do the computational and storage costs.
In either case, using Eq.~\ref{eq:lambda_cts} or Eq.~\ref{eq:lambda_arr}, a specific realization of detected counts can then be generated by Poisson sampling the mean count vector $\bvec{\lambda}$:
\begin{align}\label{eq:poisson_sample}
    \bvec{n} \sim \text{Poisson}({\bvec{\lambda}}).
\end{align}

Since Eq.~\ref{eq:lambda_arr} can also be used if the source is truly a collection of $K$ individual point sources \bvec{w}, it is our primary tool for computing expected count rates for both the array source \bvec{w} of $K$ points and a computational approximation to $w(\pvec{r})$ of $K' \gg K$ points.
Throughout this paper, we will therefore refer to both ``continuous'' and ``array'' sources as collections of discrete source points---where it is understood that the number of points $K'$ in the former will be much larger than the the number of points $K$ in the latter---and when necessary will use the prefix ``truly'' if talking about the source distributions $w(\pvec{r})$ in Eq.~\ref{eq:lambda_cts}.

It is also useful to introduce the sensitivity map $\boldsymbol{\varsigma} \in \mathbb{R}_+^K$, which is calculated for each potential source point as
\begin{align}\label{eq:sens}
    \varsigma_k = \sum_{i=1}^I \frac{\eta(\textbf{a}_i, \vec{r}_i - \pvec{r}_k) t_i}{4\pi |\vec{r}_i - \pvec{r}_k|^2}.
\end{align}
The sensitivity has dimensions of time, and thus can be interpreted as the expected number of counts per Bq of source activity at the source point $\pvec{r}_k$.

Finally, it is often useful to cast Eqs.~\ref{eq:lambda_arr} and \ref{eq:sens} in matrix form by combining all the terms in the summand except for the $w_k$ into the ``system matrix'' $\boldsymbol{V} \in \mathbb{R}_+^{I \times K}$.
Then the mean counts array is $\boldsymbol{\lambda} = \boldsymbol{V} \boldsymbol{w}$ and the sensitivity map is $\boldsymbol{\varsigma} = \boldsymbol{V}^\text{T} \boldsymbol{1}$, i.e., the sum of each column of $\boldsymbol{V}$.

\subsection{Continuous source emulation}\label{sec:emulation}
Since our primary consideration in this study is designing array sources that are relatively easy to deploy in the field, we consider only 2D regular grids of potential source locations (for both the array and continuous sources), with potentially multiple sources (i.e., variable activity) per location.\footnote{Non-planar fields and/or various support structures and fasteners could easily extend this method to~3D.}
In turn, rather than starting with an arbitrary continuous source and seeking to create a representative array source, we consider the reverse problem: given an easy-to-deploy array source, how best to create a continuous source from which it could have originated.
As we are essentially up-sampling the spatial resolution of the source, there is no unique solution to this problem.

A solution that minimizes the amount of added information is to perform a nearest-neighbor interpolation between the array points to determine the continuous point activities, and then uniformly scale the continuous source activities so that their sum matches that of the array source.
We ensure that each array point has an equal number of continuous points that are closest to it to minimize boundary effects.
As a result, we must also extrapolate this ``interpolation'' beyond the array source boundary, but only to half as many points.

\begin{figure}[!htbp]
    \centering
    \includegraphics[width=1.0\columnwidth]{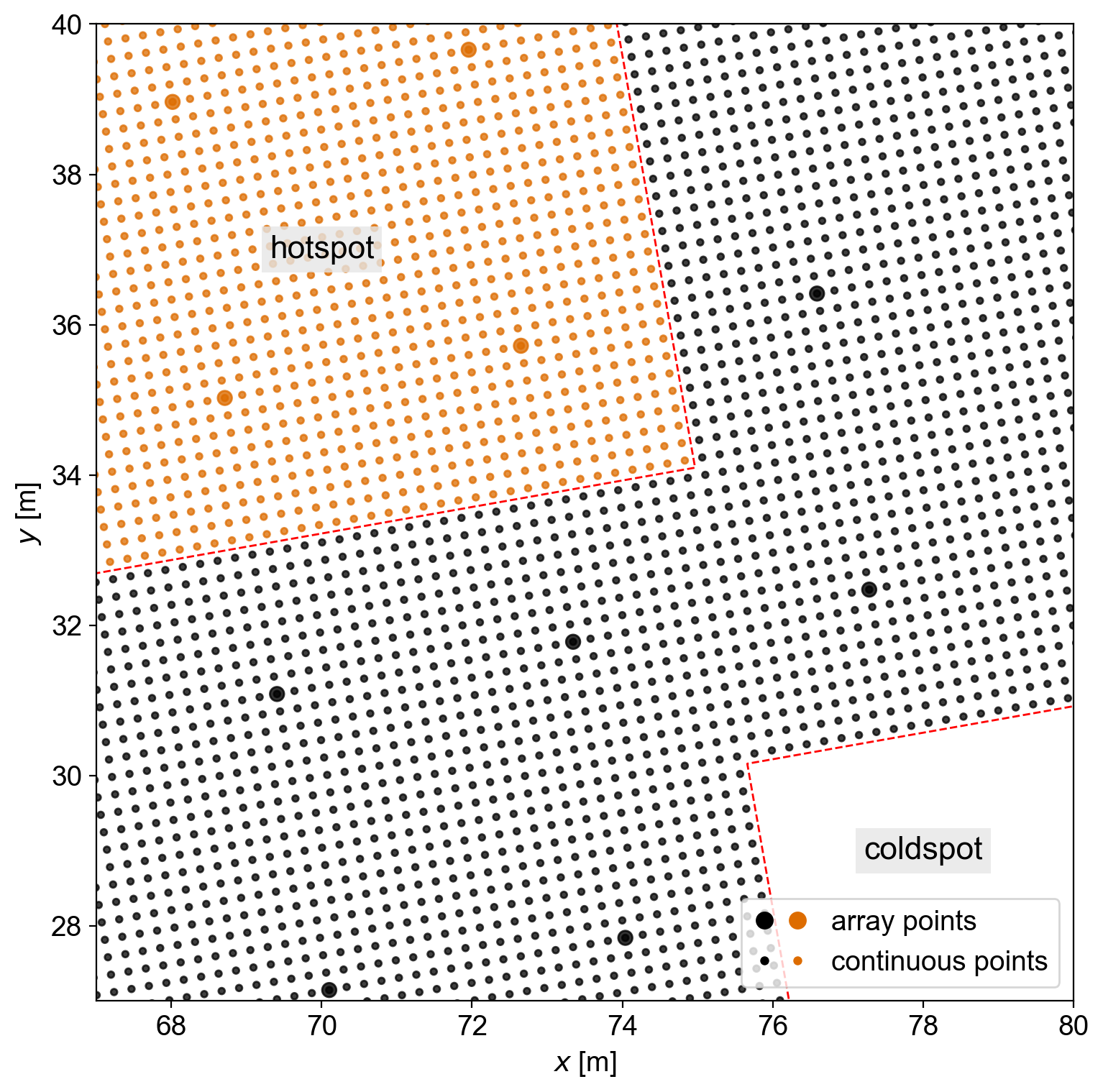}
    \caption{
        Array ($4$~m spacing) and continuous ($4\, \text{m} / 17 = 0.235$~m spacing) sources for the hot/coldspot pattern discussed in Section~\ref{sec:design}.
        The view is zoomed to more clearly show the design of the continuous source, especially at activity boundaries.
    }
    \label{fig:src_arr_vs_cts}
\end{figure}

The spacing of continuous points is chosen based on two criteria.
First, two neighboring array points must have an even number of continuous points between them, so that there is no middle point that would require activity fractionation when located at a non-uniform position within the array.
Second, the continuous spacing must be much smaller than the array spacing.
The continuous spacing required will depend on the array source spacing, as well as the detector altitude, trajectory, intrinsic efficiency, and angular resolution.
As discussed in Section~\ref{sec:design}, we designed raster pattern trajectories to cover both source and background areas of the field, with raster speeds and spacings determined by nominal UAS battery lifetimes.
Flight altitudes were limited to between $5$ and $15$~m above ground level (AGL) due to ease of operation above $5$~m and airspace restrictions above $15$~m.
The lowest flight altitudes then provide constraints on both the continuous and array source spacings.
Empirically, we find that an array spacing of $4$~m and continuous spacings of ${\lesssim}25$~cm are sufficient and computationally tractable across our parameter space given detector angular resolutions of ${\sim}10^\circ$ for $511$~keV singles for MiniPRISM and coarser for NG-LAMP.
As shown in Fig.~\ref{fig:src_arr_vs_cts}, the even-number constraint then results in $8$ continuous points on either side of an array point, with a spacing of $4\, \text{m} / (16+1) = 0.235$~m.

\subsection{Forward projections and detectors}\label{sec:forward_proj_det}

The forward projections of both the continuous and array sources to each detector (accounting for their individual position offsets) are computed at each pose of the trajectory using Eq.~\ref{eq:lambda_arr}.
The forward projection is implemented in the Python-based, GPU-accelerated {\tt mfdf} (multi-modal free-moving data fusion) library~\cite{mfdf}, which computes an array of expected counts $\boldsymbol{\lambda} \in \mathbb{R}_+^{I \times J}$, where $I$ is the number of individual measurements and $J$ is the number of individual detector elements.
This work leverages the NG-LAMP~\cite{Pavlovsky2019} and MiniPRISM~\cite{pavlovsky2019miniprism} detection systems, with $J=4$ CLLBC crystals (of size $2\times 2 \times 1$~inch) and $J=58$ CZT crystals (of size $1 \times 1 \times 1$~cm), respectively.

Detector response functions were taken from existing characterizations of the NG-LAMP and MiniPRISM detectors, which were computed ahead of time using the Geant4 framework~\cite{agostinelli2003geant4,allison2006geant4,allison2016recent}.
The detector systems (and Geant4 models thereof) include the Localization and Mapping Platform (LAMP), which consists of a LiDAR and an inertial measurement unit (IMU)---enabling LiDAR-based Simultaneous Localization and Mapping (SLAM)~\cite{hess2016real, durrant2006simultaneous, bailey2006simultaneous}---as well as a video camera, single-board computer, and front-end detector electronics---see Fig.~\ref{fig:detector_photos}.
When coupled to a UAS, the systems can also read out real-time kinematic (RTK) and standard GPS positioning measured by the UAS as a comparison for LiDAR SLAM-computed positions.

\tikzstyle{text_box}=[rectangle,fill=black!30,minimum size=1em,fill opacity=0.4,text opacity=1,text=white,font=\small\sffamily]

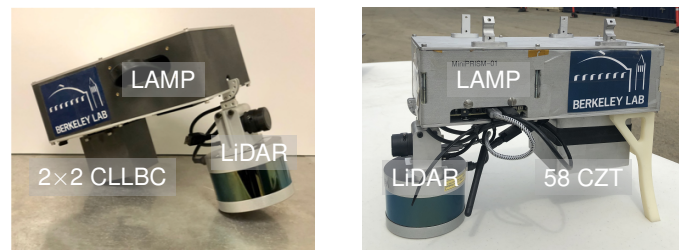
\begin{figure}[!htbp]
    \centering
    \begin{tikzpicture}
        \node[anchor=south west,inner sep=0] (image) at (0,0) {\includegraphics[height=1.27in]{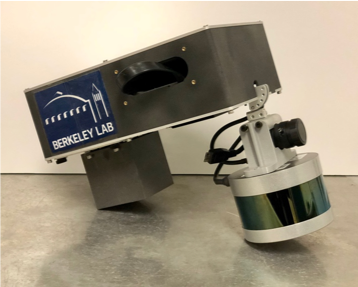}};
        \begin{scope}[x={(image.south east)},y={(image.north west)}]
            \node at (0.5,0.7) [text_box] {LAMP};
            \node at (0.8,0.4) [text_box] {LiDAR};
            \node at (0.3,0.3) [text_box] {2$\times$2~CLLBC};
        \end{scope}
    \end{tikzpicture}\hfill
    \begin{tikzpicture}
        \node[anchor=south west,inner sep=0] (image) at (0,0) {\includegraphics[height=1.27in]{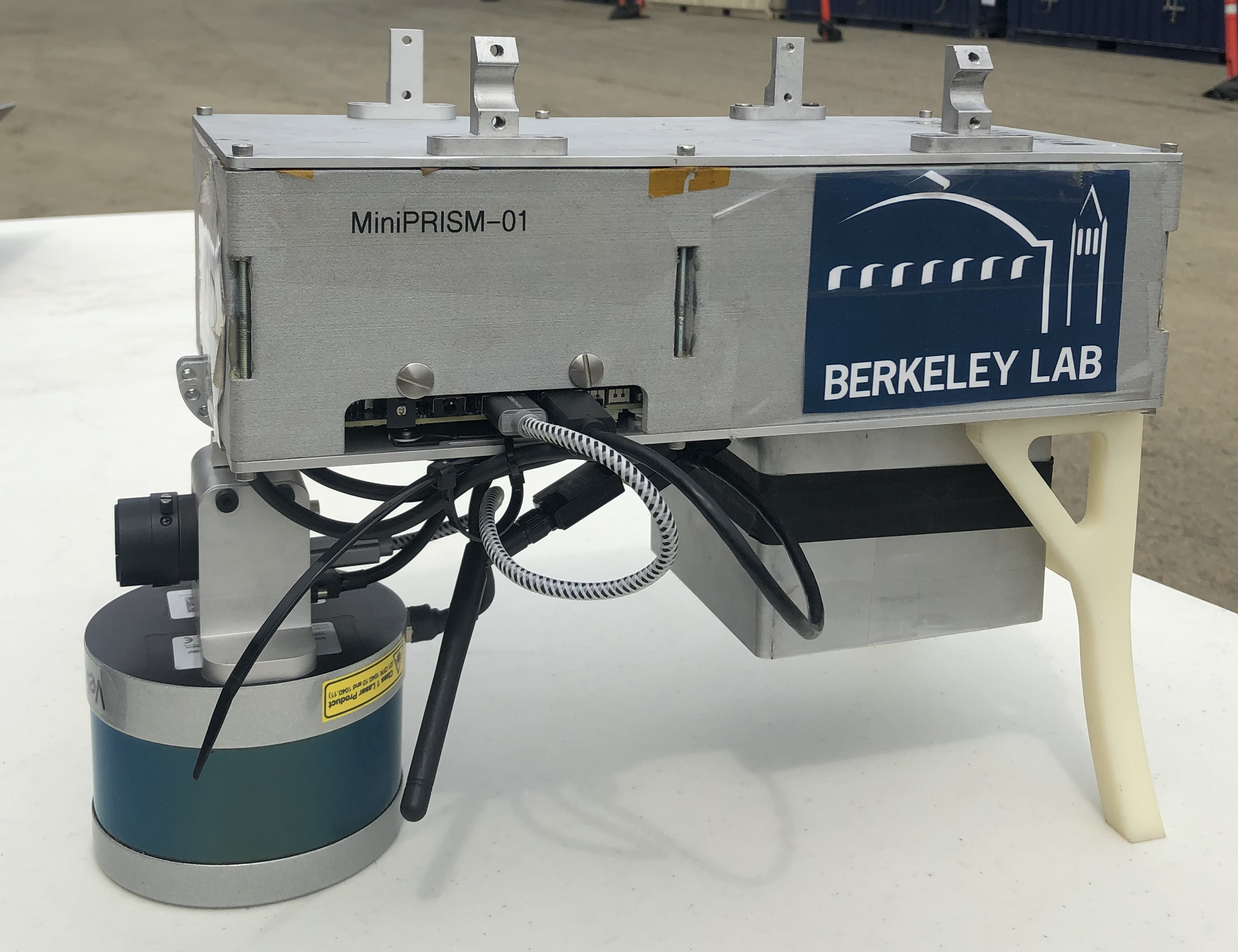}};
        \begin{scope}[x={(image.south east)},y={(image.north west)}]
            \node at (0.4,0.7) [text_box] {LAMP};
            \node at (0.2,0.3) [text_box] {LiDAR};
            \node at (0.7,0.3) [text_box] {58 CZT};
        \end{scope}
    \end{tikzpicture}
    \caption{
        Left: The NG-LAMP detector system, comprising four CLLBC crystals, a LiDAR unit, and a LAMP contextual sensor suite.
        Photo from Ref.~\cite{joshi2019experimental}.
        Right: The MiniPRISM detector system, comprising 58 CZT crystals, a LiDAR unit, and a LAMP contextual sensor suite.
        Photo from Ref.~\cite{hellfeld2021free}.
    }
    \label{fig:detector_photos}
\end{figure}

Responses were computed for both single- and double-crystal (i.e., Compton) full-energy detection efficiency at $511$~keV, though the forward projections in this work model only the single-crystal signals.
We additionally include a small but non-zero background rate in the photopeak region.
In the simulation studies discussed in Section~\ref{sec:design} we used $2$~and $0.2$~counts/s/detector for NG-LAMP and MiniPRISM, respectively, which were order-of-magnitude values estimated from previous indoor measurements.
In the experimental analysis of Section~\ref{sec:results}, however, we use $2$~counts/s after summing over detectors, as determined by dedicated handheld and aerial background measurements at the WSU field.

\subsection{Statistical tests}

To determine how well an array of point sources emulates a continuous source (for a given detector and detector trajectory), we first introduce the deviance $D(\boldsymbol{n} | \boldsymbol{\lambda})$, a scalar metric for comparing a vector of observed counts~$\boldsymbol{n}$ to a model $\boldsymbol{\lambda}$ of Poisson mean counts~\cite{baker1984clarification, mccullagh1989generalized, fisher1950significance}.
The deviance is given in ``likelihood form'' as
\begin{align}\label{eq:deviance_L}
    D(\boldsymbol{n} | \boldsymbol{\lambda}) = -2 \log L(\boldsymbol{n} | \boldsymbol{\lambda}) + 2 \log L(\boldsymbol{n} | \boldsymbol{n}),
\end{align}
where $L(\boldsymbol{n} | \boldsymbol{\lambda})$ is the Poisson likelihood of observing the data $\boldsymbol{n}$ given the model $\boldsymbol{\lambda}$, and $L(\boldsymbol{n} | \boldsymbol{n})$ is the likelihood of observing the data if the model perfectly matched the data.
Noting that $L({n_i} | \lambda_i) = e^{-\lambda_i} \lambda^{n_i} / n_i!$ for a single measurement~$i$ and that likelihoods multiply since the measurements are statistically independent, we have
\begin{align}\label{eq:deviance}
    D(\boldsymbol{n} | \boldsymbol{\lambda}) = 2 \sum_{i=1}^I \left[ n_i \log(n_i) - n_i \log(\lambda_i) + \lambda_i - n_i \right].
\end{align}
Defining
\begin{align}
    p_i &\equiv \frac{\lambda_i}{\sum_i \lambda_i} \equiv \frac{\lambda_i}{\bar{N}}\\
    q_i &\equiv \frac{n_i}{\sum_i n_i} \equiv \frac{n_i}{N}
\end{align}
we can re-write the deviance in the ``magnitude and shape form'' as
\begin{align}\label{eq:deviance_magshape}
    D(\boldsymbol{n} | \boldsymbol{\lambda}) = D(N | \bar{N}) + 2 N \DKL{\boldsymbol{q}}{\boldsymbol{p}},
\end{align}
where $D(N | \bar{N})$ is the deviance between the observed sum of counts $N$ and the model sum of counts $\bar{N}$, and $\DKL{\boldsymbol{q}}{\boldsymbol{p}}$ is the Kullback–Leibler divergence between the normalized count vectors $\boldsymbol{q}$ and $\boldsymbol{p}$.
Note that in the case of multiple detector elements, $\boldsymbol{q}$ and $\boldsymbol{p}$ are formed from concatenating the $\boldsymbol{q}$ and $\boldsymbol{p}$ of each separate detector.

For multiple Poisson noise realizations of the $I$ measurements, the distribution of the deviance statistic can be approximated by a shifted Gamma distribution whose first three moments match the calculated moments of the deviance statistic.
The Gamma distribution is chosen to extend the moment-matching technique from a symmetric Gaussian distribution to one with a non-zero skew term.
The parameters of this shifted Gamma distribution are expensive to calculate as the number of measurements $I$ gets large, but two limiting cases exist when the number of counts $n_i$ in most measurements is ${\gtrsim}30$:
\begin{enumerate}
    \item As the gross counts $N$ become large, the deviance distribution can be well-approximated by a $\chi^2$ distribution with $I$ degrees of freedom.
    \item As the number of measurements $I$ becomes large, the $\chi^2$ distribution itself can be well-approximated by a normal distribution with mean $I$ and variance $2I$.
\end{enumerate}
These limits will often not apply due to the low counts per measurement far from the source, so in general we will use the Gamma distributions.
As shown in the later Fig.~\ref{fig:deviances}, however, the deviance distributions are often still well-described by normal distributions, but with means and variances different from $I$ and $2I$.

We can now define the model $\boldsymbol{\lambda}_0$ as the mean count vector for our continuous source, and the model $\boldsymbol{\lambda}_1$ for our coarsely-gridded array source.
Measurements $\boldsymbol{n}_0$ and $\boldsymbol{n}_1$ of the continuous and array sources will produce deviances of $\Dnl{0}{0}$ and $\Dnl{1}{1}$, respectively.
(Note that in the aforementioned large $N$ and large $I$ limits, both $\Dnl{0}{0}$ and $\Dnl{1}{1}$ will converge to the same distribution.)

We now know in principle what the deviance distributions will look like, given that we know which is the true model for a particular measurement, $\boldsymbol{\lambda}_0$ or $\boldsymbol{\lambda}_1$.
We denote their probability density functions (pdfs) as $\mathbb{P}(D | \boldsymbol{\lambda}_0)$ and $\mathbb{P}(D | \boldsymbol{\lambda}_1)$ for true models $\boldsymbol{\lambda}_0$ and $\boldsymbol{\lambda}_1$, respectively.
If we do not know the true model, however, we can ask what the deviance distribution will look like when the model is mis-specified.
In particular, we can ask how the deviances $\Dnl{1}{0}$ are distributed when the data are generated from an array source ($\boldsymbol{n}_1$) but deviances are calculated assuming a continuous source ($\boldsymbol{\lambda}_0$).

We can then compute the theoretical shifted Gamma distributions of the deviances assuming the data samples $\boldsymbol{n}_0$ and $\boldsymbol{n}_1$ are Poisson samples from $\boldsymbol{\lambda}_0$ and $\boldsymbol{\lambda}_1$.
In particular we compute four theoretical deviance distributions for each parameter combination (see for instance the later Fig.~\ref{fig:devs_src0_fp0_nglamp}):
\begin{enumerate}
    \item $\Dnl{0}{0}$: the deviance distribution when Poisson samples $\boldsymbol{n}_0$ are generated from the continuous source model~$\boldsymbol{\lambda}_0$
    \item $\Dnl{1}{1}$: the deviance distribution when Poisson samples $\boldsymbol{n}_1$ are generated from the array source model~$\boldsymbol{\lambda}_1$
    \item $\Dnl{1}{0}$: the deviance distribution when Poisson samples $\boldsymbol{n}_1$ are generated from the array source model~$\boldsymbol{\lambda}_1$ but their deviances are calculated assuming the continuous model $\boldsymbol{\lambda}_0$ is the correct model.
    \item $\Dll{1}{0}$: a constant cross term that originates from assuming the incorrect model; this is not a true deviance as it does not compare Poisson counts to a model, but it does have the same functional form.
\end{enumerate}
The shifted Gamma distribution of the ``non-central'' deviances $\Dnl{1}{0}$ is computed from the first three moments in essentially the same fashion as the central deviances.
Namely, since each deviance statistic is the sum of statistically independent terms, the mean, variance, and third central moment of the deviance are simply the sums of those same moments for the terms.
Those three moments are numerically estimated using the relevant Poisson distribution.
For example, following Eq.~\ref{eq:deviance}, the $i\textsuperscript{th}$ term of $\Dnl{1}{0}$ is
\begin{align}
    2 (n_i \log n_i - n_i \log \lambda_{0i} + \lambda_{0i} - n_i)
\end{align}
and the mean, variance, and third central moment of this term can be estimated assuming $n_i \sim \mathrm{Poisson}(\lambda_{1i})$.
In the aforementioned large-$N$ and large-$I$ limits, these calculations also allow one to compute the theoretical parameters of the $\chi^2$ or Gaussian approximations for $\Dnl{1}{0}$.

Equipped with this framework, we can now describe how well the array source mimics the continuous source.
Quantitatively, given the continuous and array models $\boldsymbol{\lambda}_0$ and $\boldsymbol{\lambda}_1$, what is the probability that a sample $\boldsymbol{n}_1$ drawn from $\boldsymbol{\lambda}_1$ ``looks like'' it was drawn from $\boldsymbol{\lambda}_0$?
I.e., what is the false negative probability $P_\text{FN}$ for incorrectly deciding that the most likely model $\hat{\boldsymbol{\lambda}}$ is the continuous source $\boldsymbol{\lambda}_0$ when the true model is the array source $\boldsymbol{\lambda}_1$?
In the absence of prior information, the decision rule for choosing the most likely model $\hat{\boldsymbol{\lambda}}$ for an observed deviance $D$ is simply
\begin{align}
    \hat{\boldsymbol{\lambda}} =
    \begin{cases}
    \boldsymbol{\lambda}_0,\quad D \leq D^\star\\
    \boldsymbol{\lambda}_1,\quad D \geq D^\star
    \end{cases}
\end{align}
where the decision threshold $D^\star$ is the value of $D$ at which both models are equally likely:
\begin{align}
    D^\star = \underset{D}{\text{argwhere}}\: \mathbb{P}(D | \boldsymbol{\lambda}_0) = \mathbb{P}(D | \boldsymbol{\lambda}_1).
\end{align}
Then the false negative probability is
\begin{align}
    P_\text{FN} &= \mathbb{P}(\hat{\boldsymbol{\lambda}} = \boldsymbol{\lambda}_0 | \boldsymbol{\lambda} = \boldsymbol{\lambda}_1)\\
    &= \mathbb{P}(D < D^\star | \boldsymbol{\lambda} = \boldsymbol{\lambda}_1)\\
    &= \text{CDF}_{\boldsymbol{\lambda}_1}(D^\star),
\end{align}
where $\text{CDF}_{\boldsymbol{\lambda}_1}$ is the cumulative distribution function of $D$ given $\boldsymbol{\lambda} = \boldsymbol{\lambda}_1$.
As expected, the probability that a sample $\boldsymbol{n}_1$ drawn from the array source $\boldsymbol{\lambda}_1$ ``looks like'' it was drawn from a continuous source $\boldsymbol{\lambda}_0$---and thus the degree to which the array source can be used as a useful proxy of a continuous source---depends on the overlap of the two deviance pdfs.
Intuitively, a perfect ``spoof'' should have $P_\text{FN} = 1/2$---it is indistinguishable from the continuous source via the deviance metric.
In this work, we relax this perfect spoof condition and consider an array source to be a \textit{practical} spoof of a continuous source if it has $0.4 \leq P_\text{FN} \leq 0.5$, though we note that this lower bound is somewhat arbitrary.

\subsection{Example calculations}\label{sec:example_calcs}
\begin{figure}[!htbp]
    \centering
    \includegraphics[width=1.0\columnwidth]{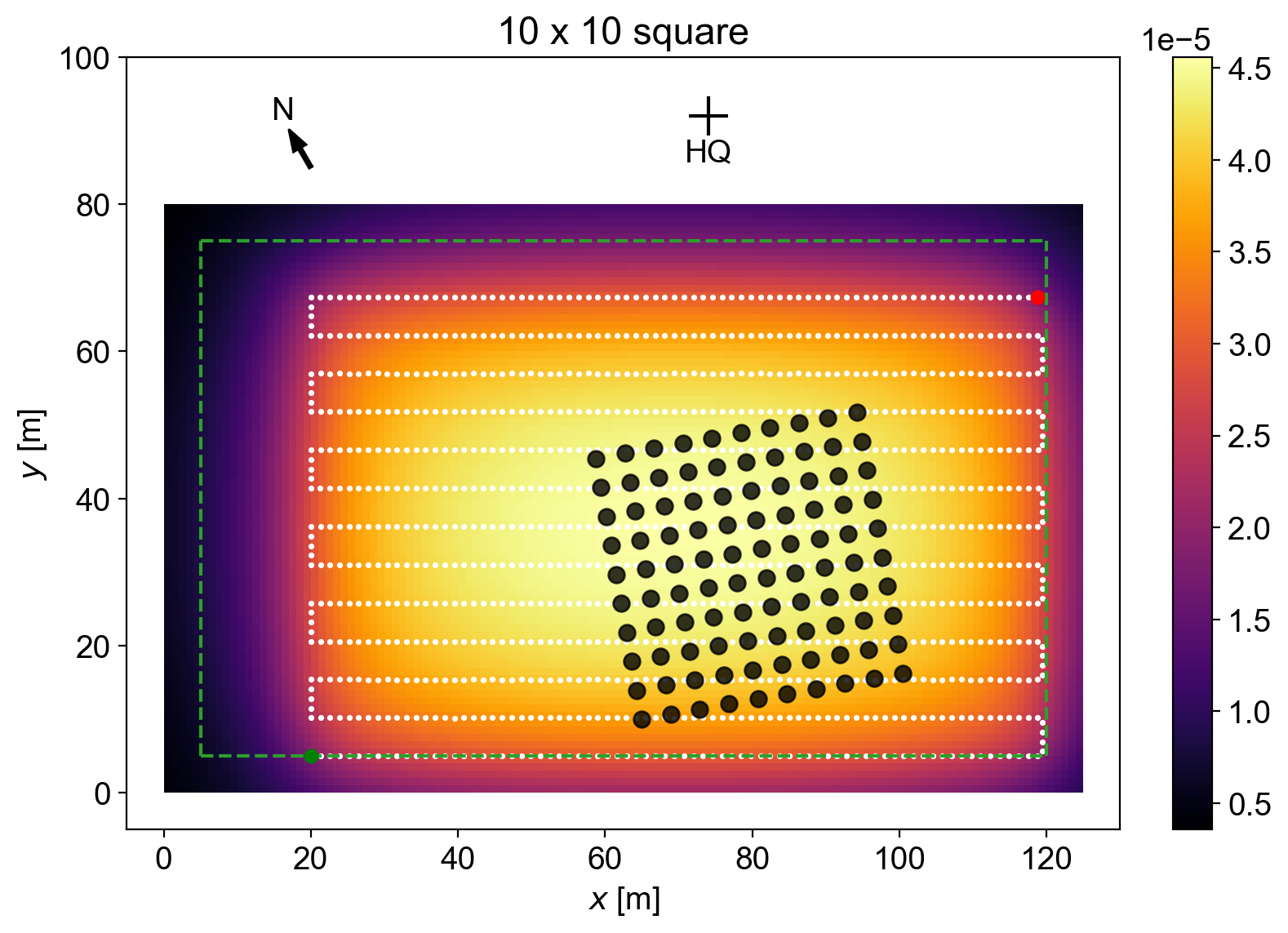}
    \caption{
        Layout of the synthetic $10 \times 10$ square source (black points) and a typical $100$~m-wide synthetic UAS raster pattern (dotted white line) with $13$ lines spaced at $5.2$~m.
        The start and stop of the trajectory are indicated by green and red circles, respectively.
        The green dashed rectangle denotes a $5$~m flight buffer from the field boundary.
        The approximate compass direction and HQ location are also denoted.
        The sensitivity map $\boldsymbol{\varsigma}$ at $6$~m AGL is shown beneath the source points and ranges from ${\sim}0.5$--$4.5 \times 10^{-5}$~counts/Bq over the field extent.
        Over the source extent, the dense raster spacing maintains a relatively uniform sensitivity of ${\sim}3.5$--$4.5 \times 10^{-5}$~counts/Bq.
    }
    \label{fig:src_config_0}
\end{figure}

\begin{figure}[!htbp]
    \centering
    \includegraphics[width=1.0\columnwidth]{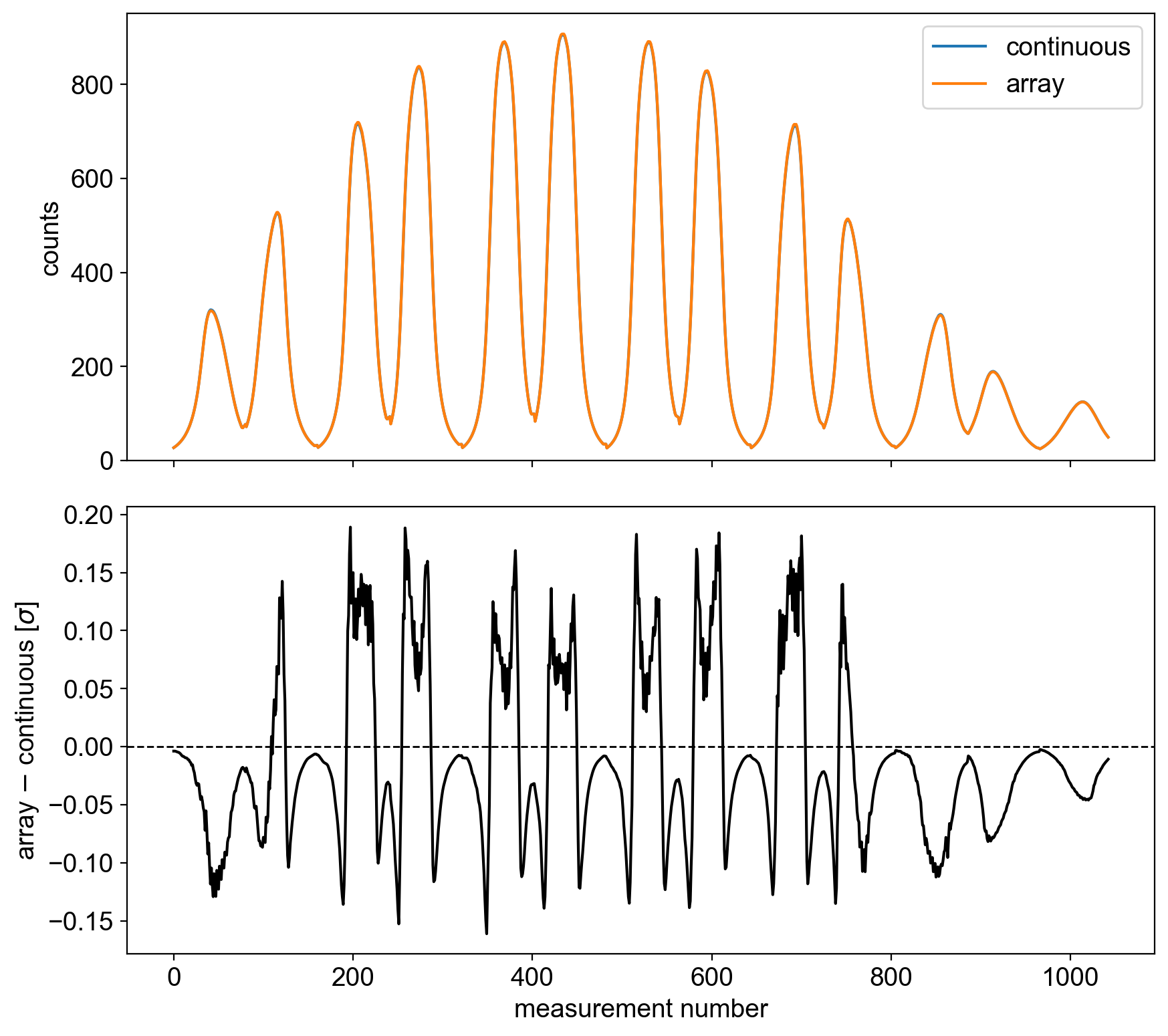}
    \caption{
        Top: forward projections of the synthetic mean counts of the continuous source ($\boldsymbol{\lambda}_0$) and the array source ($\boldsymbol{\lambda}_1$) for the scenario in Fig.~\ref{fig:src_config_0}.
        The ``continuous" curve in blue is almost completely obscured by the ``array" curve in orange.
        Bottom: comparison of the two $\boldsymbol{\lambda}$, showing statistical differences (in this case, generally below $0.2\sigma$ absolute value) in the array source (and its Poisson error) compared to the continuous source.
        Counts are binned to $t_i = 0.5$~s and summed over the $J=4$ NG-LAMP detector elements.
    }
    \label{fig:counts_cts_vs_arr}
\end{figure}

\begin{figure}[!htbp]
    \centering
    \includegraphics[width=1.0\columnwidth]{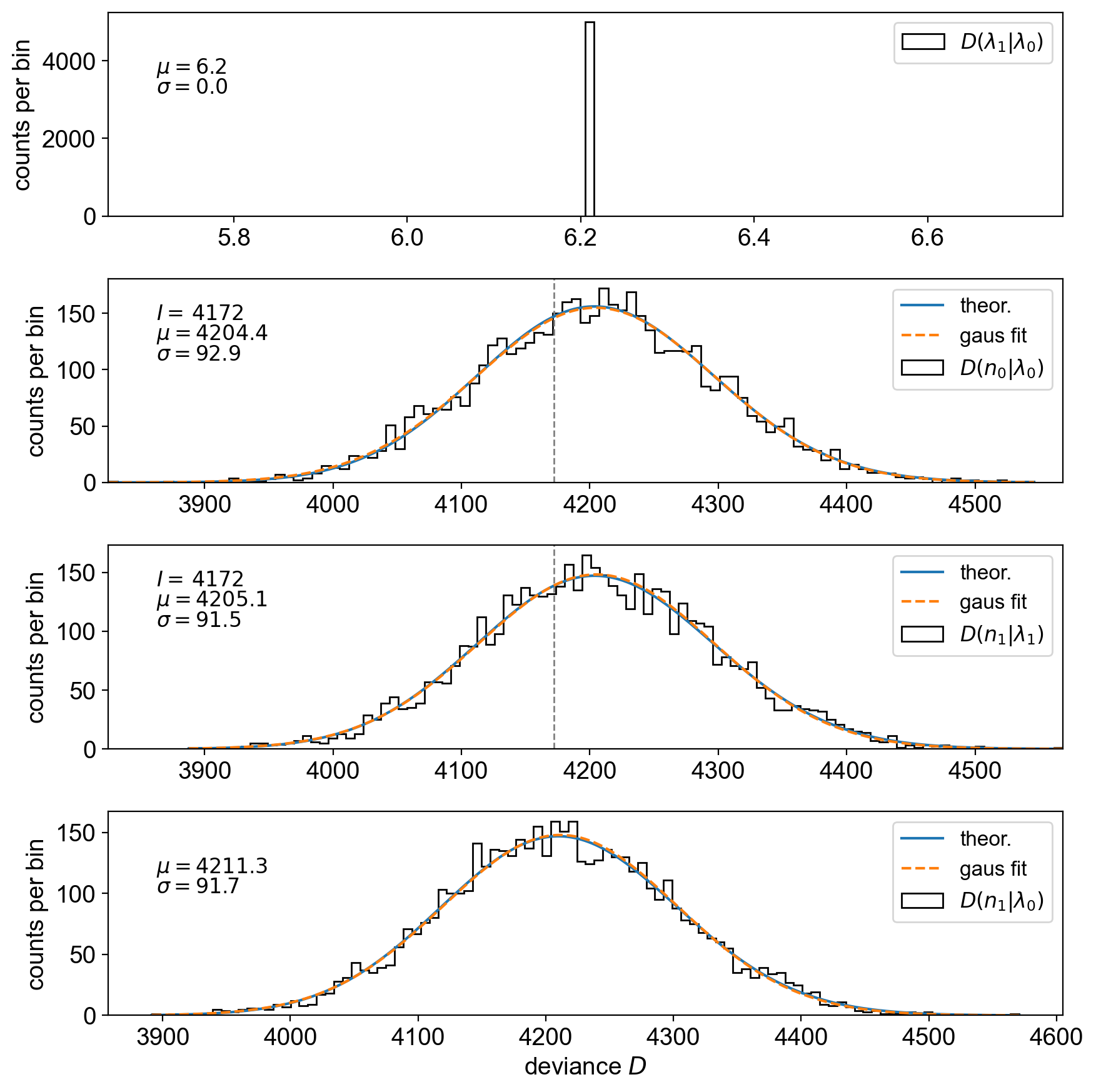}
    \caption{
        Deviance distributions for the NG-LAMP raster pattern over the $10 \times 10$ square of $5$~mCi sources shown in Fig.~\ref{fig:src_config_0}.
        The histograms show the result of Poisson sampling~$\boldsymbol{\lambda}$ $5000$ times, while the solid and dashed curves show the theoretical shifted Gamma distributions and Gaussian fits to the histogram, respectively.
        In the top plot, the standard deviation $\sigma$ is $0$ since only the two mean count arrays are compared.
        Since only Poisson noise is present, the $\Dll{1}{0}$ is constant.
        In the middle two plots, a dashed vertical line is drawn at $D = I$, the number of measurements, showing that distributions are approximately but not exactly centered at $I$.
        Top: $\Dll{1}{0}$ between the two models.
        Top middle: $\Dnl{0}{0}$ for the continuous source.
        Bottom middle: $\Dnl{1}{1}$ for the array source.
        Bottom: $\Dnl{1}{0}$ for the array source compared to the continuous source.
    }
    \label{fig:devs_src0_fp0_nglamp}
\end{figure}

\begin{figure}[!htbp]
    \centering
    \includegraphics[width=1.0\columnwidth]{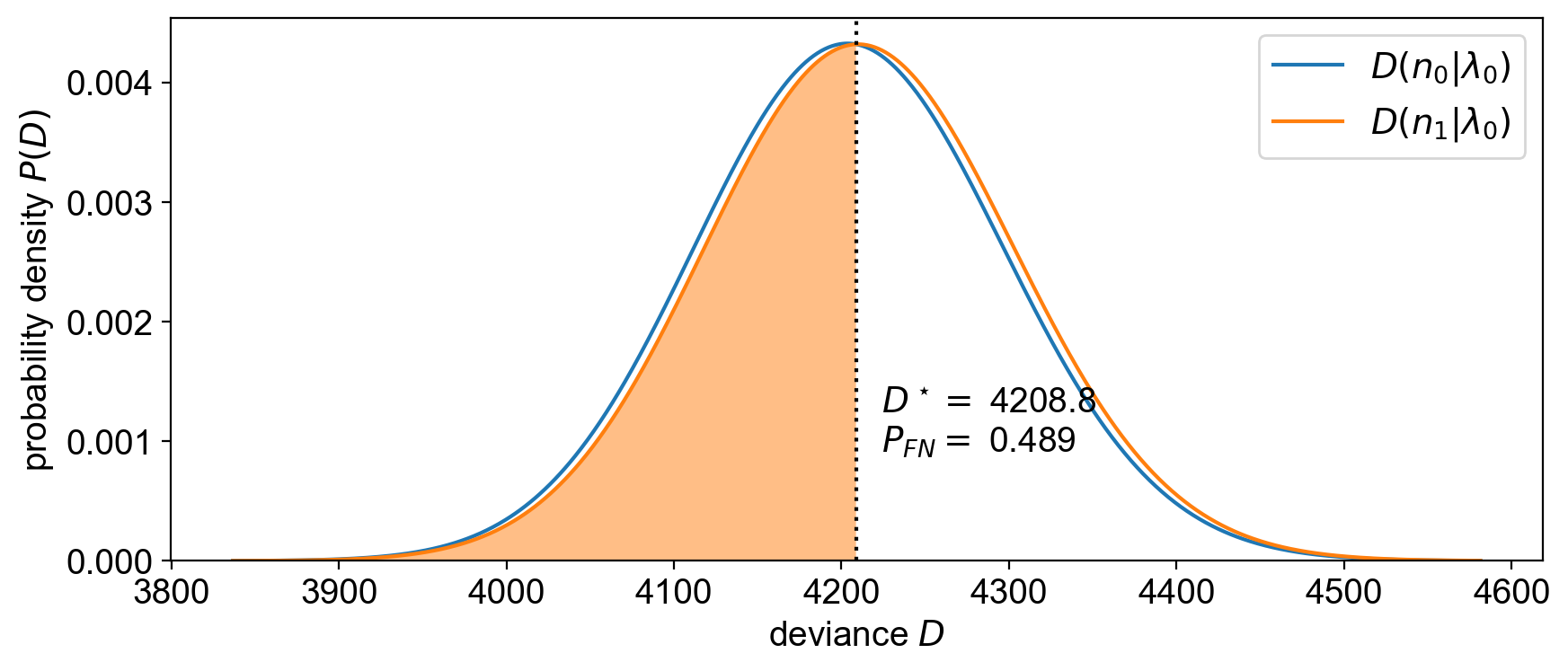}
    \caption{
        Deviances from Fig.~\ref{fig:devs_src0_fp0_nglamp} used to determine the false negative probability $P_\text{FN} = 0.489$.
    }
    \label{fig:deviances}
\end{figure}

Fig.~\ref{fig:src_config_0} shows an example scenario consisting of a $10 \times 10$ square array source pattern and a $6$~m (AGL) NG-LAMP raster trajectory---see Section~\ref{sec:design} for additional information on the design of the source and raster patterns.
Fig.~\ref{fig:counts_cts_vs_arr} shows for this scenario the expected counts measured by the NG-LAMP detector with both the array ($\boldsymbol{\lambda}_1$) and continuous ($\boldsymbol{\lambda}_0$) sources.
Here $I = 1043$~measurements with a time binning of $t_i = 0.5$~s.
Fig.~\ref{fig:devs_src0_fp0_nglamp} shows the theoretical and empirical distributions of deviance statistics computed from the two $\boldsymbol{\lambda}$ arrays after $5000$ Monte Carlo samples, and Fig.~\ref{fig:deviances} shows the comparison of the deviance distributions $\Dnl{0}{0}$ and $\Dnl{1}{0}$ used for computing the false negative probability~$P_\text{FN}$.

Fig.~\ref{fig:src_config_0} also serves to introduce the Cartesian ``field coordinate system'' used throughout this work.
The origin is placed at the southwest corner of the field, and $x$ and $y$ are measured along the fence lines to the southeast and northwest corners, respectively.
We note the positive $y$ direction differs from north by roughly $30^\circ$.
The $z$ direction therefore defines elevation above the ground level, which is assumed to be perfectly flat.

\section{Experimental design}\label{sec:design}

\begin{figure}[!htbp]
    \centering
    \includegraphics[width=0.49\columnwidth]{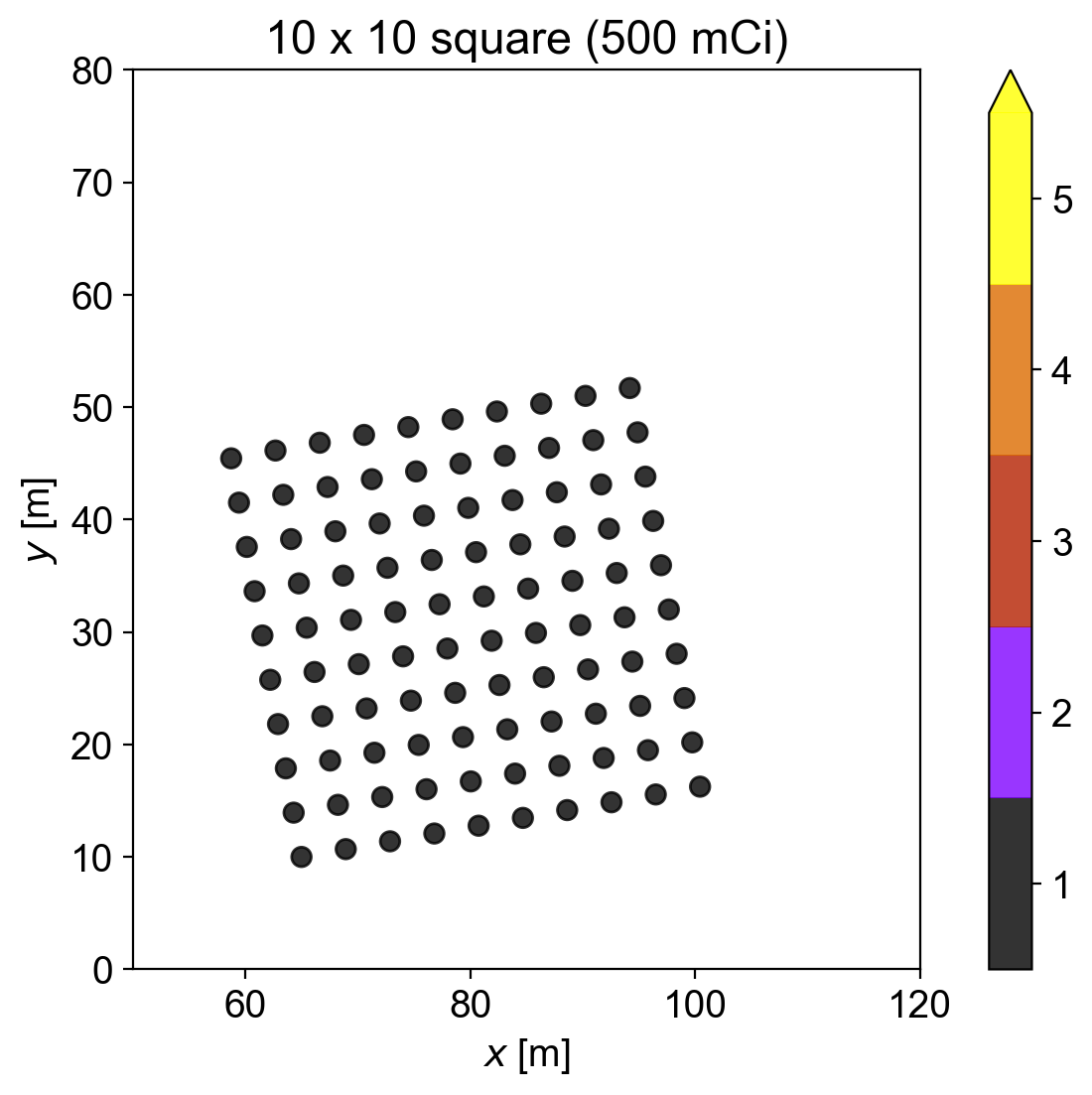}\hfill
    \includegraphics[width=0.49\columnwidth]{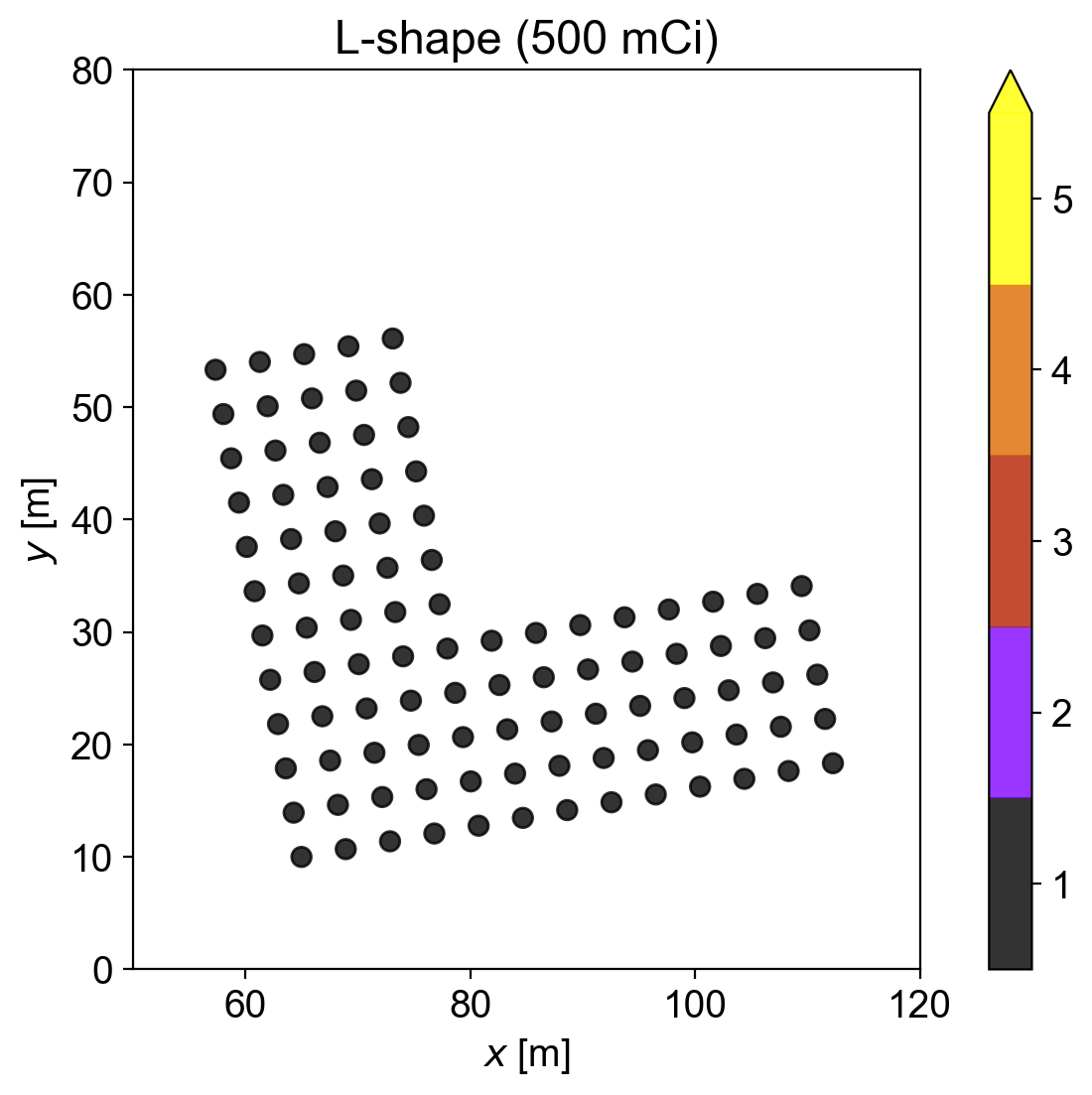}\\
    \includegraphics[width=0.49\columnwidth]{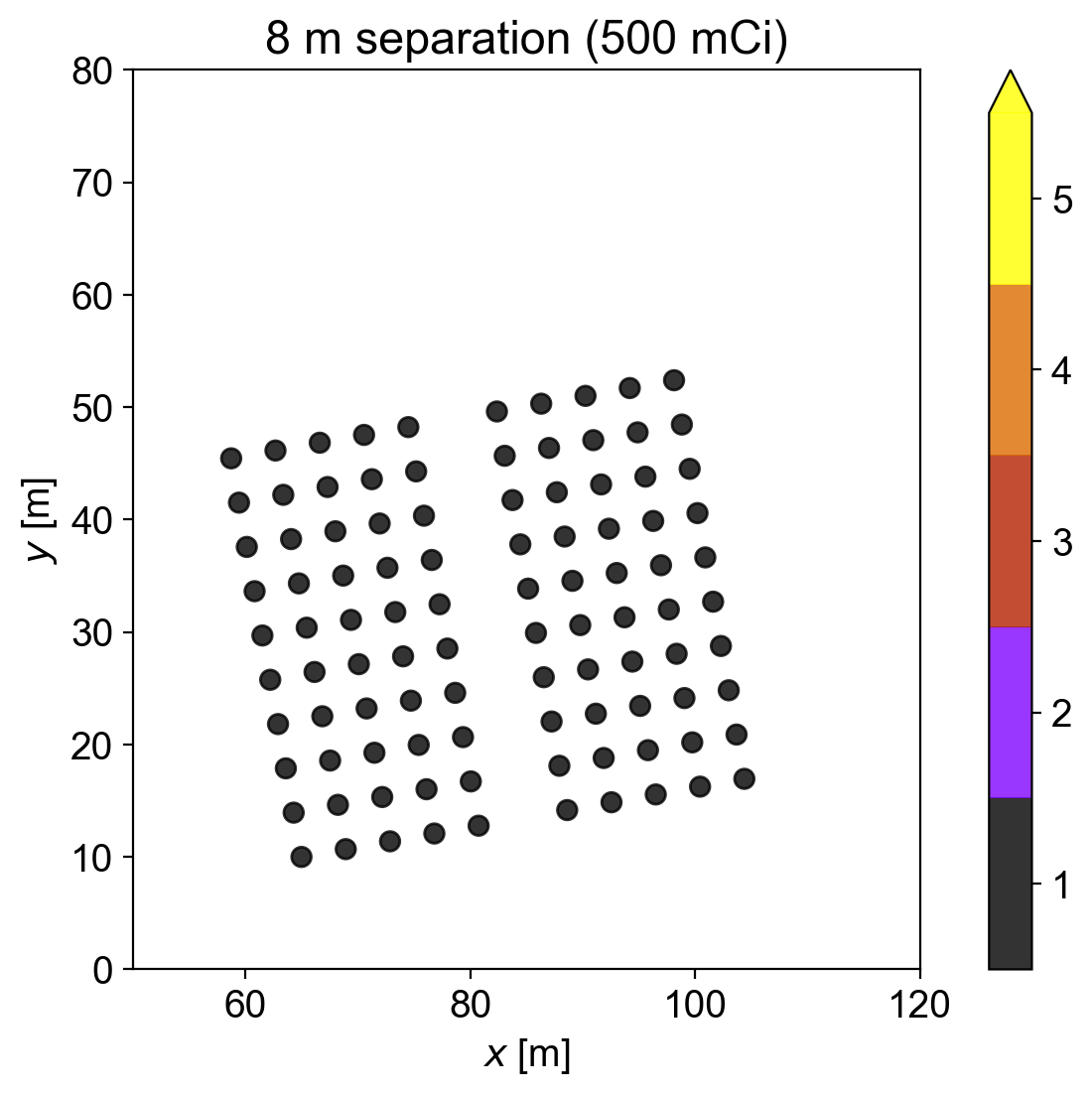}\hfill
    \includegraphics[width=0.49\columnwidth]{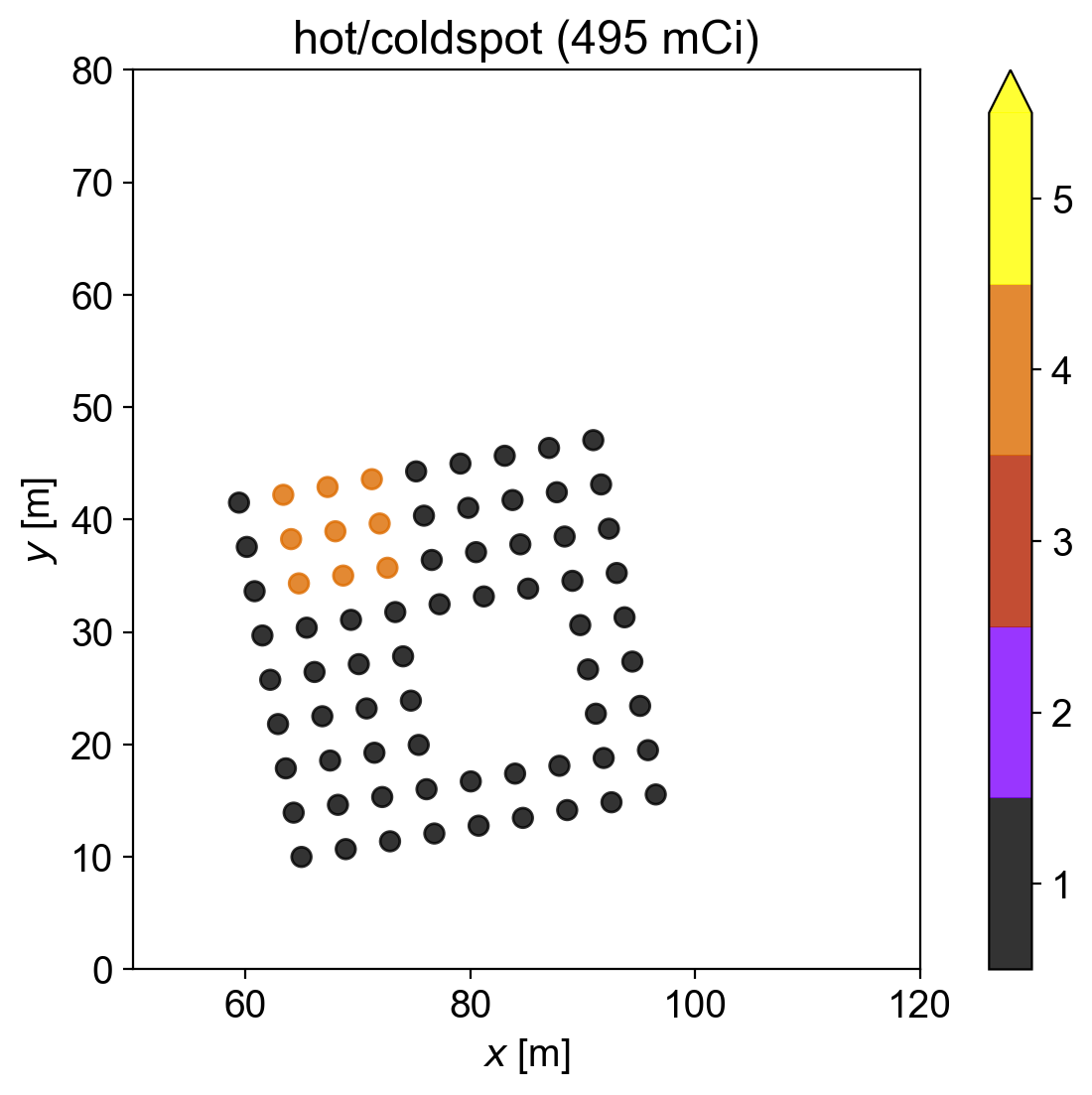}\\
    \includegraphics[width=0.49\columnwidth]{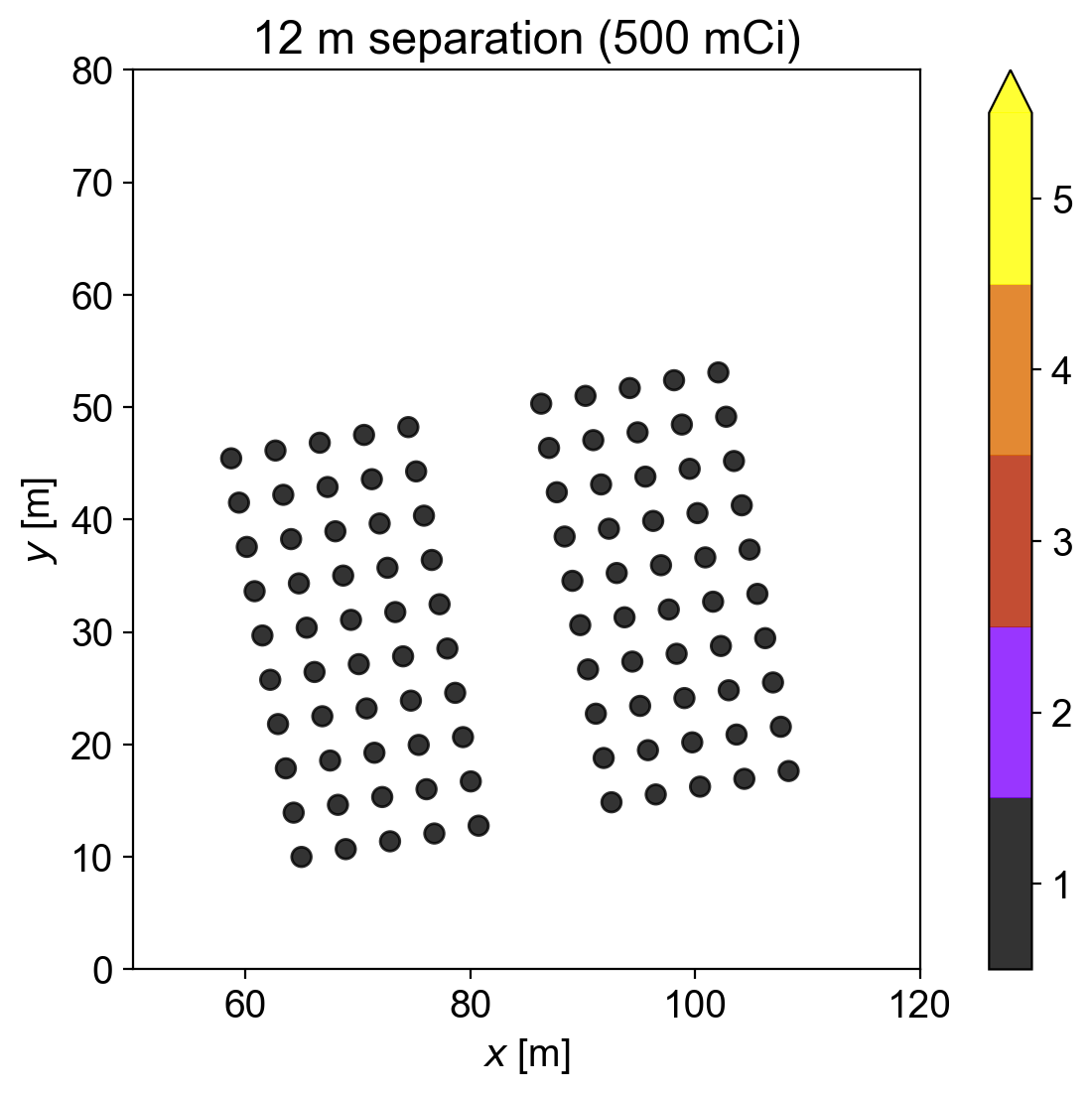}\hfill
    \includegraphics[width=0.49\columnwidth]{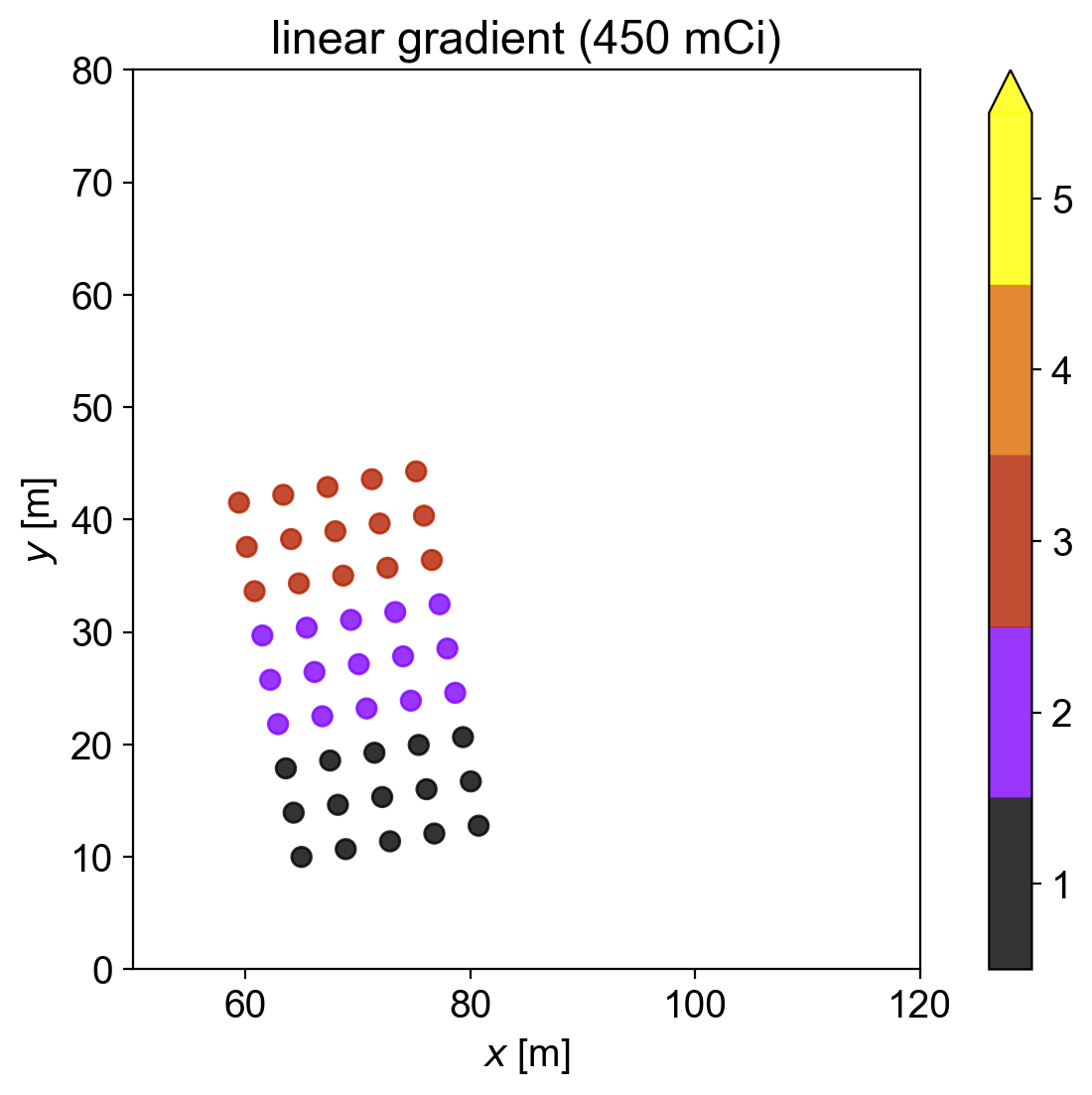}\\
    \includegraphics[width=0.49\columnwidth]{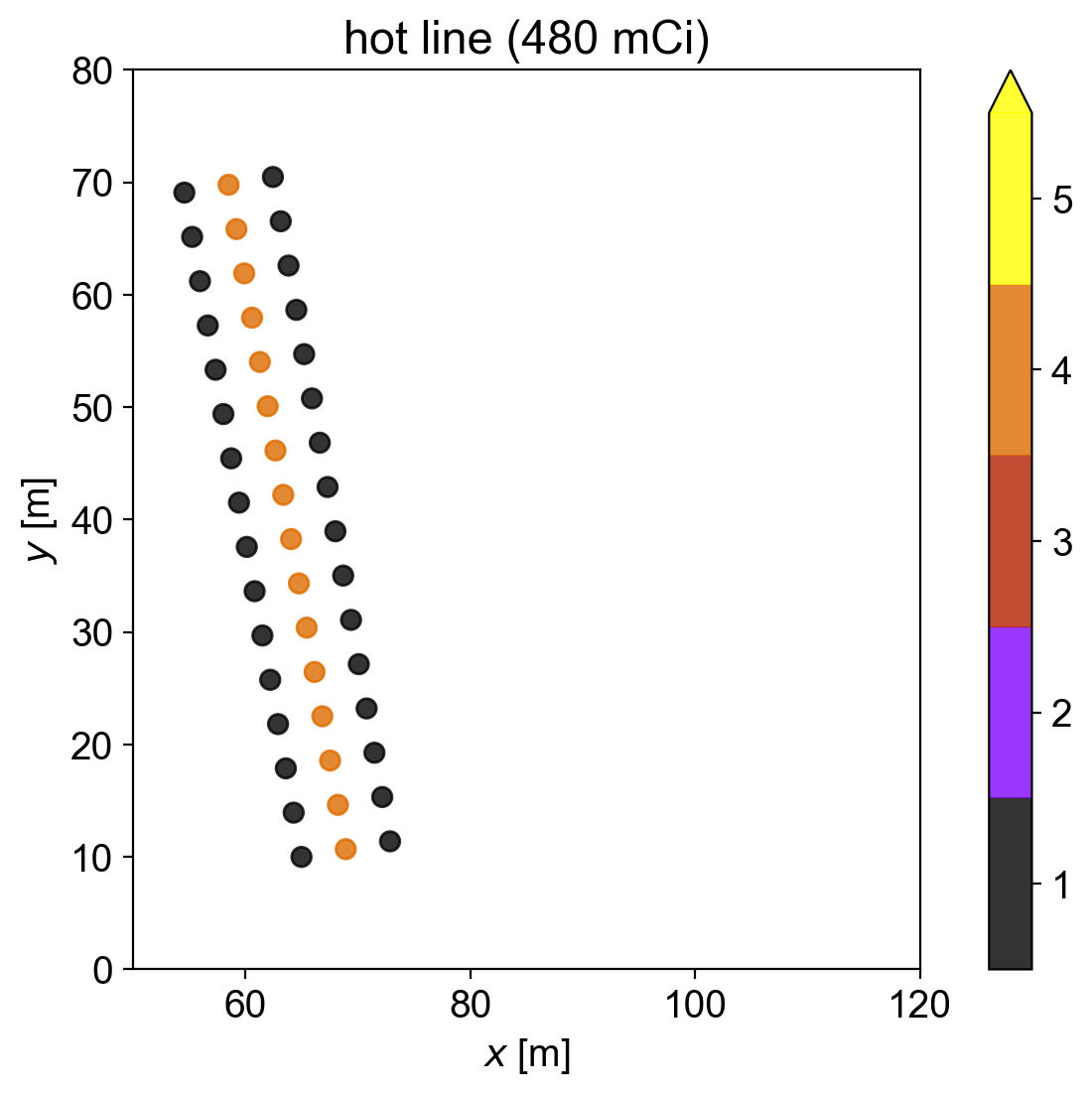}\hfill
    \includegraphics[width=0.49\columnwidth]{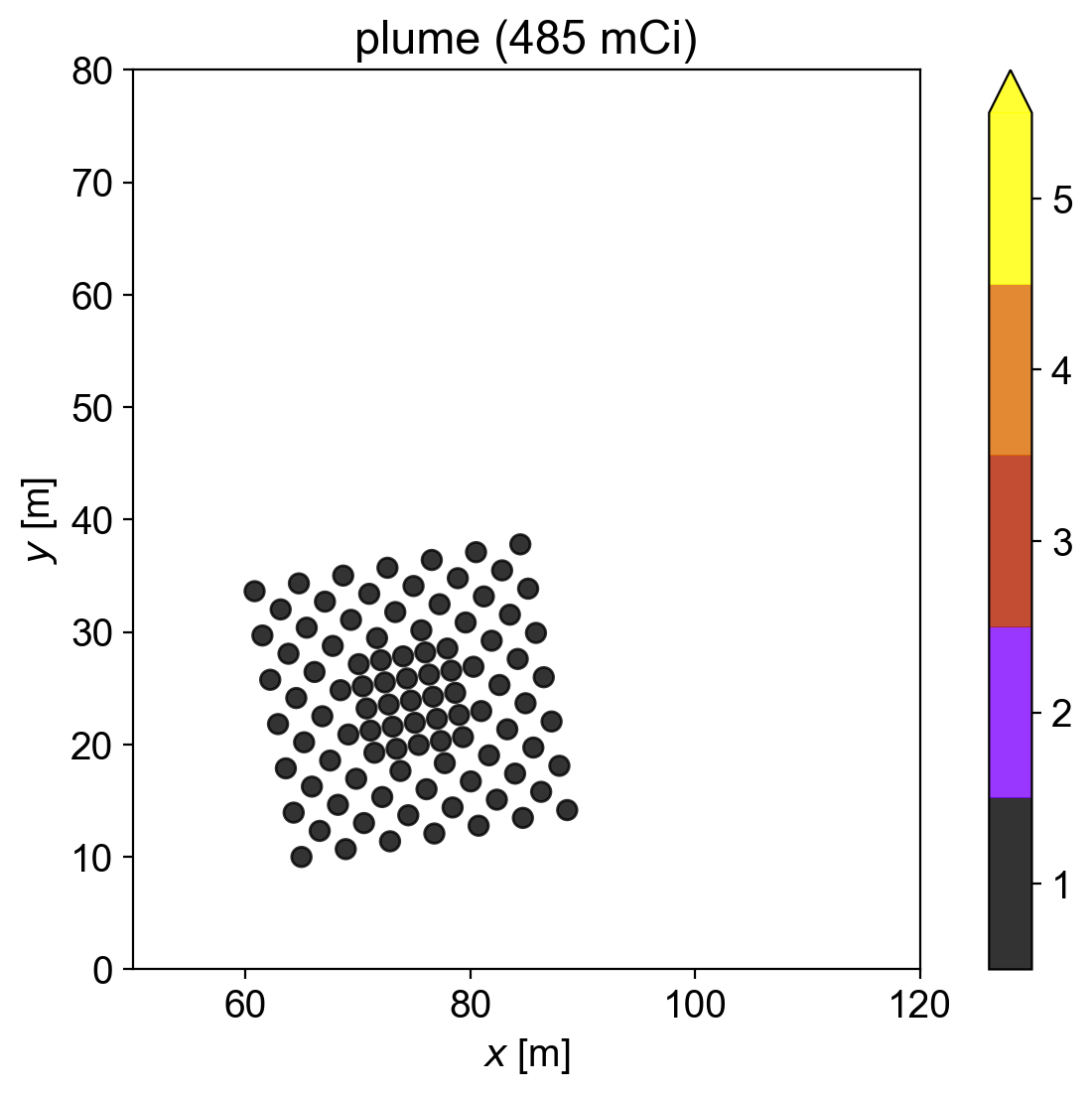}
    \caption{
        All eight source configurations deployed during the WSU measurement campaign.
        The color bar denotes the number of nominal $5$~mCi Cu-64 sources at each point.
    }
    \label{fig:all_patterns}
\end{figure}

Eight planar array sources consisting of up to $100$ individual $5$~mCi Cu-64 sources each were designed for use during the measurement campaign (see Fig.~\ref{fig:all_patterns}).
These configurations comprise:
\begin{enumerate}[beginpenalty=10000]
    \item a $10 \times 10$ square;
    \item a $5 \times 9$ rectangle with a gradient in intensity that is produced by three adjacent $5\times3$ rectangles containing three, two and one source per point;
    \item a $9 \times 9$ square with inset $3 \times 3$ grids of four sources per point and no sources, referred to as the ``hot/coldspot'';
    \item a pair of $5 \times 10$ rectangles separated by two grid spacings ($8$~m);
    \item a pair of $5 \times 10$ rectangles separated by three grid spacings ($12$~m);
    \item an \Lshape{} with $12\,(13)$ sources on its short (long) dimension and a thickness of five sources;
    \item a $3 \times 16$ line with a hot center, which has four sources per point, referred to as the ``hot line''; and
    \item a ``plume'' consisting of a $7 \times 7$ outer checkerboard pattern and a $5 \times 5$ fully-occupied central region.
\end{enumerate}
As shown in Fig.~\ref{fig:all_patterns}, with the exception of the plume source, the source spacing in all source patterns is $4$~m.
The $4$~m separation was chosen to provide good $P_\text{FN}$ values while also creating spatially large source distributions.
We note that while calculations were performed for $5$~mCi sources, the sources decayed non-negligibly throughout the day with $t_{1/2} = 12.7$~hours, and were found to be closer to ${\sim}7$~mCi at the start of each measurement day---see Table~\ref{tab:source_info} and Section~\ref{sec:discussion}.

The sources share a common lower-left corner at $(x, y) = (65, 10)$~m, chosen to reduce dose rates to personnel at the headquarters (HQ) and create large empty regions on the west side of the field (see Fig.~\ref{fig:src_config_0}) to enhance contrast between source and background regions.
Each array source pattern is rotated $10^\circ$ about its lower-left corner to reduce the effect of aliasing with the UAS raster pattern, which is aligned with the field boundaries.

The above sources were designed for a variety of measurement goals, the analysis for which will be covered in a later work.
The $10 \times 10$ square, being symmetric and uniform, provides a simple baseline with which to test reconstruction quality (via metrics such as the Structural Similarity Index Metric (SSIM), root-mean-squared error (RMSE), total activity, uniformity of activity, and edge sharpness) as well as repeatability over multiple measurements.
The \Lshape{} was similarly designed as a uniform source with an interior corner.
We note that a modified version of our \Lshape{} could provide an array source analogue to the truly continuous source in Ref.~\cite{chen2020flight}.
The separated pairs of $5 \times 10$ rectangles were designed as uniform sources with narrow corridors of zero activity; such patterns can be analyzed for how well-separated the two rectangles are after reconstruction as a function of measurement parameters (e.g., altitude, detector, and the rectangle separation itself) and provides opportunities for studying dose-minimizing or information-maximizing path-planning algorithms.
The linear gradient source was designed as a simple non-uniform source with which to test the reconstructed activity dropoff.
Similarly, the hot/coldspot source provides non-uniform activities as well as internal regions of zero activity; this source will be used to again test the ``sharpness'' of the reconstruction.
The hot line source was designed as a high-contrast source and also to test the resolution for narrow shapes with a spatial gradient.
Finally, the plume source was also designed to test the activity change between the outer and inner regions, while also testing denser source placements.

The trajectories flown by the UAS were also designed with a number of competing goals in mind.
First, a dense field-aligned raster pattern was chosen to reduce the parameter space, avoid obstacles adjacent to the field, generate an approximately uniform sensitivity in the source region, and provide opportunities for studies of sparser raster patterns (e.g., only every second raster line) by cutting measurements in post-processing rather than by retaking data.
As discussed in Section~\ref{sec:emulation}, flight altitudes were limited to between $5$ and $15$~m.
The UAS orientation was course-aligned in order to improve the LiDAR coverage vs.\ a fixed orientation.
The raster spacing of $5.2$~m, length of $100$~m, and speed of $2.6$~m/s were chosen to both overfly the source extent and collect data over zero-source regions for improved background estimation while completing in ${\lesssim}10$~min to make the most use of the battery life.
Originally, the raster was designed to traverse the entire long dimension of the field (up to the $5$~m buffer on each side), but was shortened during the measurement campaign to approximately the $100$~m width shown in Fig.~\ref{fig:src_config_0} to account for lower-than-expected UAS battery performance.
Similarly, most raster patterns were started from the bottom left corner of the field so that the passes over the source would be completed first in case of an early landing.
Moreover, it was found that that this trajectory typically led to acceptable false negative probabilities of $0.4 \leq P_\text{FN} \leq 0.5$, and to generally accurate MAP-EM reconstructions of the simulated source shape and intensity from the forward-projected $\boldsymbol{\lambda}$.
We hypothesize this agreement between the simulated true and reconstructed sources is due in large part to the relatively high and uniform sensitivity (Eq.~\ref{eq:sens}) over the true source extent afforded by this trajectory across various altitudes, including the $6$~m altitude AGL used for most NG-LAMP flights---see again Fig.~\ref{fig:src_config_0}.
In particular, using an altitude larger than the pass spacing tends to smooth out variations in the sensitivity due to the large constant $z$ term in the distance between the detector and a given source voxel.

\section{Source fabrication and assay}\label{sec:fabrication_and_assay}

\subsection{Source fabrication}\label{sec:fabrication}
Cu-64 was chosen as an attractive nuclide for the distributed sources measurement campaign for several reasons.
First, the $\beta^{+}$ decay of Cu-64 to Ni-64 results in a prominent annihilation photon line at $511$~keV with a yield of $0.352$ photons per disintegration~\cite{singh2007nuclear, qaim2007positron}.
This energy is suitable for both singles and Compton imaging, and is close to the $662$~keV line from long-lived Cs-137 contamination following reactor accidents such as those at Chernobyl and Fukushima~\cite{imanaka2015comparison}.
Second, while not strongly interfering with measurements of $511$~keV photons, the weak $1346$~keV line provides a convenient check on the analysis, albeit at very limited statistics.
Third, high-elemental-purity ($99.999$--$99.9999\%$) copper pellets (natural abundances $69.15\%$~Cu-63, $30.85\%$~Cu-65) were obtainable from American Elements~\cite{ae_copper}, reducing impurities produced during the neutron irradiation of the pellets.
These high-purity pellets could also be massed to achieve one pellet per source in order to reduce handling requirements and simplify source tracking.
Solid metal pellets also reduced the risk of accidental radionuclide release that would be present with powdered or liquid sources, especially in an outdoor setting.
Finally, the short half-lives of the neutron capture products Cu-64 ($t_{1/2}=12.7$~hours) and Cu-66 ($t_{1/2}=5.10$~minutes) ensured that relatively large total activities could be produced without any long-lived radioactive waste.

\begin{figure}[!htbp]
    \centering
    \includegraphics[
        width=0.49\columnwidth,
        trim={20cm, 20cm, 40cm, 10cm},  
        clip,
        keepaspectratio
    ]{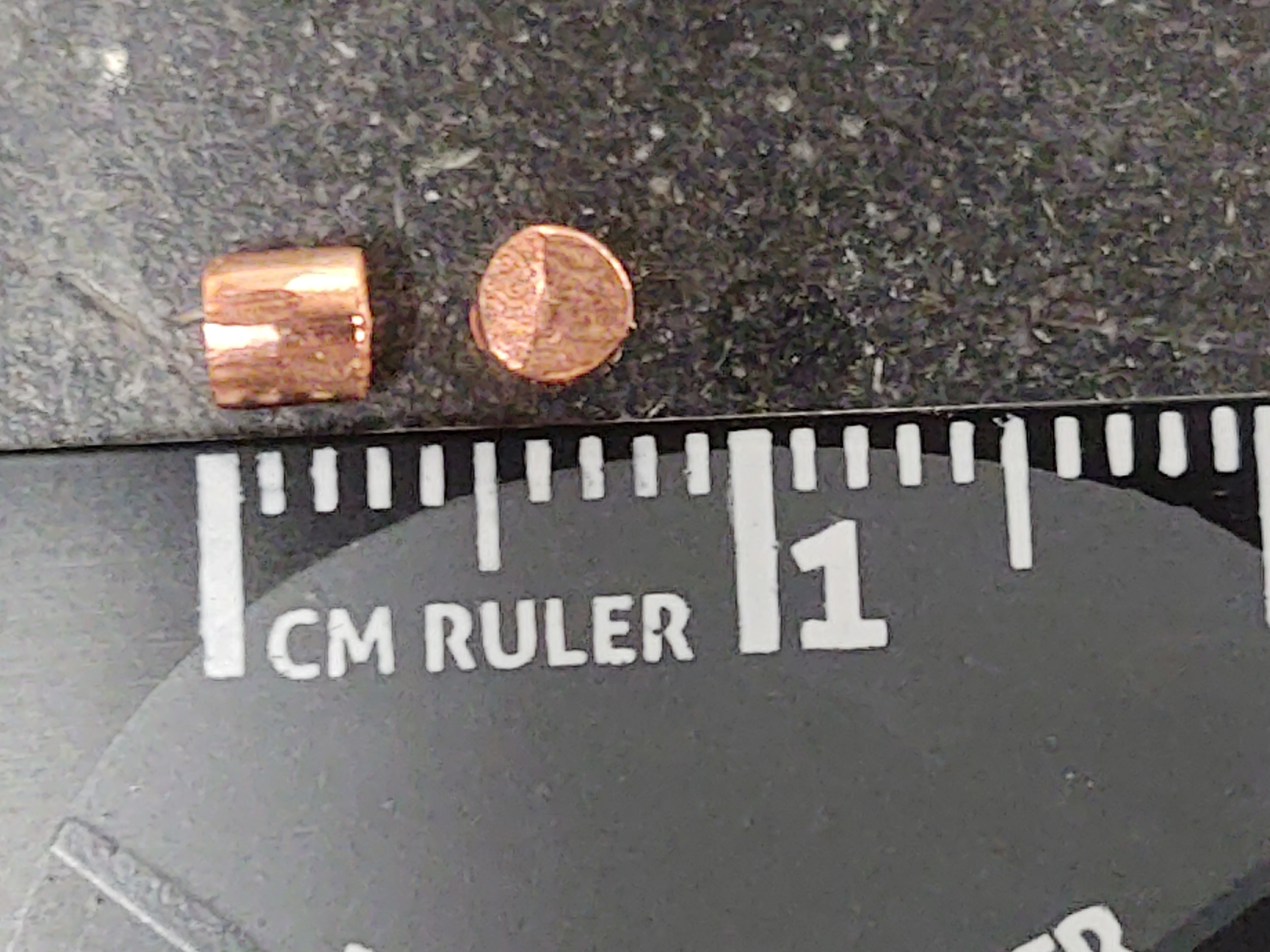}\hfill
    \includegraphics[
        width=0.49\columnwidth,
        trim={20cm, 15cm, 40cm, 15cm},
        clip,
        keepaspectratio
    ]{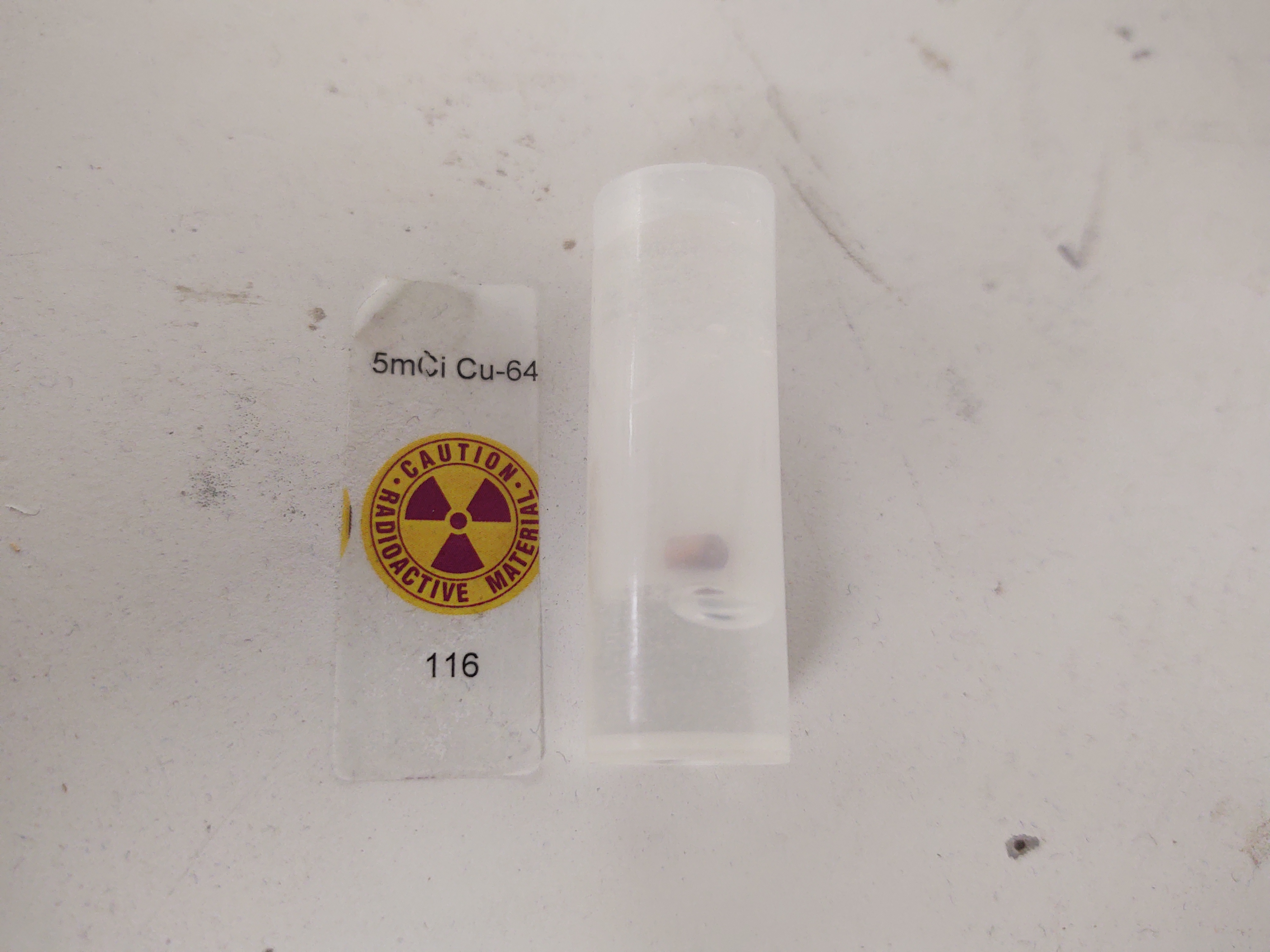}
    \caption{
        Left: two representative copper pellets.
        Right: a copper pellet encapsulated in epoxy and double-encased in vials.
    }
    \label{fig:copper_pellets}
\end{figure}

To produce the Cu-64 sources, three batches of $100$ high-purity copper pellets were irradiated for up to $2400$~s by the thermal and epithermal neutron flux of the WSU Nuclear Science Center $1$~MW TRIGA reactor.
Following irradiation, the copper pellets were stored in the reactor pool; approximately six hours after irradiation, each pellet was transferred into a 2/5-dram vial that was pre-epoxied into a 2-dram vial (see Fig.~\ref{fig:copper_pellets}).
Once the copper pellet was transferred, the remainders of its 2/5- and 2-dram vials were filled with epoxy.
After each $100$-source batch was finished and distributed into epoxied vials, the sources were left to cure overnight.
The next morning the 2-dram vials containing the sources were capped, heat sealed, and checked for contamination.
Contamination swipes found no removable contamination present.
The sealed and cured material was loaded into tennis balls with an opening slit cut into them for transport and use on the field.
Tennis balls were chosen for their ease of both visually tracking on the field during the exercises and ease of manipulation and replacement by long-handled grabber tools.
In total, the sources were allowed to cool for approximately a day between irradiation and use on the field, allowing the Cu-66 component ($t_{1/2}=5.10$~minutes) to completely decay out.

\subsection{Source assay}\label{sec:assay}
Following the distributed sources measurement campaign, the Cu-64 sources were assayed via high-purity germanium (HPGe) gamma spectroscopy to determine the ground truth pellet activities.
In the initial post-experiment gamma assay, however, the epoxy embedding of the pellets caused large, uncontrolled variations in the source-to-detector distance compared to the source positions of the available calibration standards.
These distance variations caused the assays to measure $50\%$ relative activity variations across samples that were irradiated in very similar conditions and were expected to have very nearly uniform activity. 
As a result, $18$ representative copper pellets were re-irradiated under reactor conditions as similar as possible to the first irradiations, with minimal changes to the neutron flux due to fuel burn-up in the interim.
After cooling for $165$~hours, the $18$ pellets---this time not encased in epoxy---were individually surveyed by HPGe detectors.
During the $300$~s assays, ${\sim}2 \times 10^5$ net counts were recorded in the $511$~keV peak per pellet, with dead times of ${\sim}5\%$.

The activities assayed in this second (henceforth ``September'') batch of pellets were then used to determine the expected activities produced by the first (henceforth ``August'') set of irradiations.
In particular, the irradiation process can be described by a neutron point-kinetics model for the Cu-64 population in a copper pellet as a function of time.
We define $\phi(t)$ as the one-group mean apparent neutron flux across the volume of the copper pellet, rather than the true incident flux on the boundary of the copper cylinder.
As such, we do not need to explicitly account for correction factors such as self-shielding~\cite{lindstrom2008neutron}.
The Cu-64 number density \Nfourt{} during irradiation is then governed by the differential equation\footnote{The neutron-induced destruction of Cu-64 is negligible.
Extending Eq.~\ref{eq:point_kinetics} to include such a term effectively modifies the destruction coefficient $\lambda_{64} \to \lambda_{64} + \sigma_D \phi$, where $\sigma_D$ is the Cu-64 destruction cross section.
The full $\sigma_D$ is unknown, but the radiative capture component has been measured to be $270 \pm 170$~b~\cite{kneff1986helium}.
Given a thermal flux of ${\sim}5\times 10^{16}$~neutrons/m$^2$/s, the resulting change in destruction rate is ${\sim}0.01\%$.
}
\begin{align}\label{eq:point_kinetics}
    \frac{\dd \Nfourt}{\dd t} = \phi(t) \Nthreet{} \sigma_C - \lambda_{64} \Nfourt,
\end{align}
where \Nthree{} is the Cu-63 number density, $\sigma_C \simeq 4.5$~b~\cite{zerkin2018experimental} is the (thermal group) Cu-63 $\to$ Cu-64 radiative capture cross section, and $\lambda_{64}$ is the Cu-64 decay constant.
The activity $\Afourt{}$ associated with this number density is
\begin{align}\label{eq:act_and_ndens}
    \Afourt{} = \Nfourt{} \lambda_{64} m_\text{pellet} / \rho_\text{Cu},
\end{align}
where $m_\text{pellet}$ is the mass of the given copper pellet and $\rho_\text{Cu}$ is the density of copper.
Assuming that the flux is constant in time $\phi(t) \equiv \phi$, that the change in the Cu-63 population is negligible $\Nthreet{} \equiv \Nthree$, and that the initial Cu-64 population is $N_{64}(0) = 0$, we have
\begin{align}\label{eq:sol_point_kinetics}
    \Nfourt = \frac{\phi \Nthree{} \sigma_C}{\lambda_{64}} \left[ 1 - e^{-\lambda_{64}t} \right].
\end{align}
We note that Cu-63 term $\Nthree{}$ can be further expanded in terms of the Cu-63 natural abundance ratio $f_{63}$, the molar mass of copper $A_\text{Cu}$, and Avogadro's number $N_\text{Av}$ as
\begin{align}
    \Nthree{} = f_{63} \rho_\text{Cu} N_\text{Av} / A_\text{Cu}.
\end{align}

Given irradiation (``cook'') and decay (``cool'') time durations \Dtcook{} and \Dtcool{}, respectively, we can write
\begin{align}\label{eq:sol_post_cool}
    \Nfour(t_f) = \frac{\phi \Nthree{} \sigma_C}{\lambda_{64}} \left[ 1 - e^{-\lambda_{64} \Dtcook} \right] e^{-\lambda_{64} \Dtcool}.
\end{align}
where $t_f$ denotes the ``final'' or ``field'' time at the end of the cooling window and thus the start of the September HPGe assay or August UAS measurement day.
We can then rearrange Eq.~\ref{eq:sol_post_cool} for the apparent neutron flux
\begin{align}\label{eq:flux_apparent}
    \phi = \frac{\Nfour(t_f)}{\Nthree} \frac{\lambda_{64}}{\sigma_C} \left[ 1 - e^{-\lambda_{64} \Dtcook} \right]^{-1} \left[ e^{-\lambda_{64} \Dtcool} \right]^{-1}.
\end{align}
Eqs.~\ref{eq:sol_post_cool} and~\ref{eq:flux_apparent} thus form a pair of forward and backward models that can be used to determine the August pellet activities based on the September assay data.
In Eq.~\ref{eq:flux_apparent}, the activity $\Afour(t_f)$---and thus number density $\Nfour(t_f)$ via Eq.~\ref{eq:act_and_ndens}---is determined from the (net) number of Cu-64 counts $C_{64}$ observed during an HPGe spectroscopy measurement, using either the $511$~keV or $1346$~keV spectral line.
To first order (i.e., assuming the assay livetime $\Delta t_\text{assay}$ is short compared to the Cu-64 lifetime), we can write
\begin{align}\label{eq:counts_and_act}
    C_{64} &= \Afour(t_f) \Delta t_\text{assay} \epsilon b\\
           &= \Nfour(t_f) V_\text{pellet} \lambda_{64} \Delta t_\text{assay} \epsilon b
\end{align}
where $\epsilon$ is the total detection efficiency (at either $511$~keV or $1346$~keV, depending on which region of interest is used to define the counts $C_{64}$), and $b$ is the branching ratio or gammas per decay again depending on the spectral line assayed.
We note that while the approximate Eq.~\ref{eq:counts_and_act} is shown for clarity of presentation, our analyses correct for the slight (${\sim}1\%$) decay of the source activity during the measurement.
The net number of counts $C_{64}$ in each measurement is determined by integrating either the $511$~keV or $1346$~keV peak region of interest (ROI) and subtracting a small constant background based on the ${\sim}5$~keV windows on either side of the ROI.
As discussed further in Section~\ref{sec:discussion}, the $511$~keV annihilation peak is broader than the HPGe resolution for a nuclear decay line at the same energy, and thus requires a broader ROI to accurately determine the net counts.
Once the $C_{64}$ are determined, the mean apparent neutron flux $\phi$ can then be computed from the September irradiations via Eq.~\ref{eq:flux_apparent}, and substituted back into Eq.~\ref{eq:sol_post_cool} with the August \Dtcook{} and \Dtcool{} to determine the expected August pellet number densities $\Nfourt{}$ and thus activities at the start of each UAS measurement day in Table~\ref{tab:source_info}.

\begin{figure}[!htbp]
    \centering
    \includegraphics[width=1.0\columnwidth]{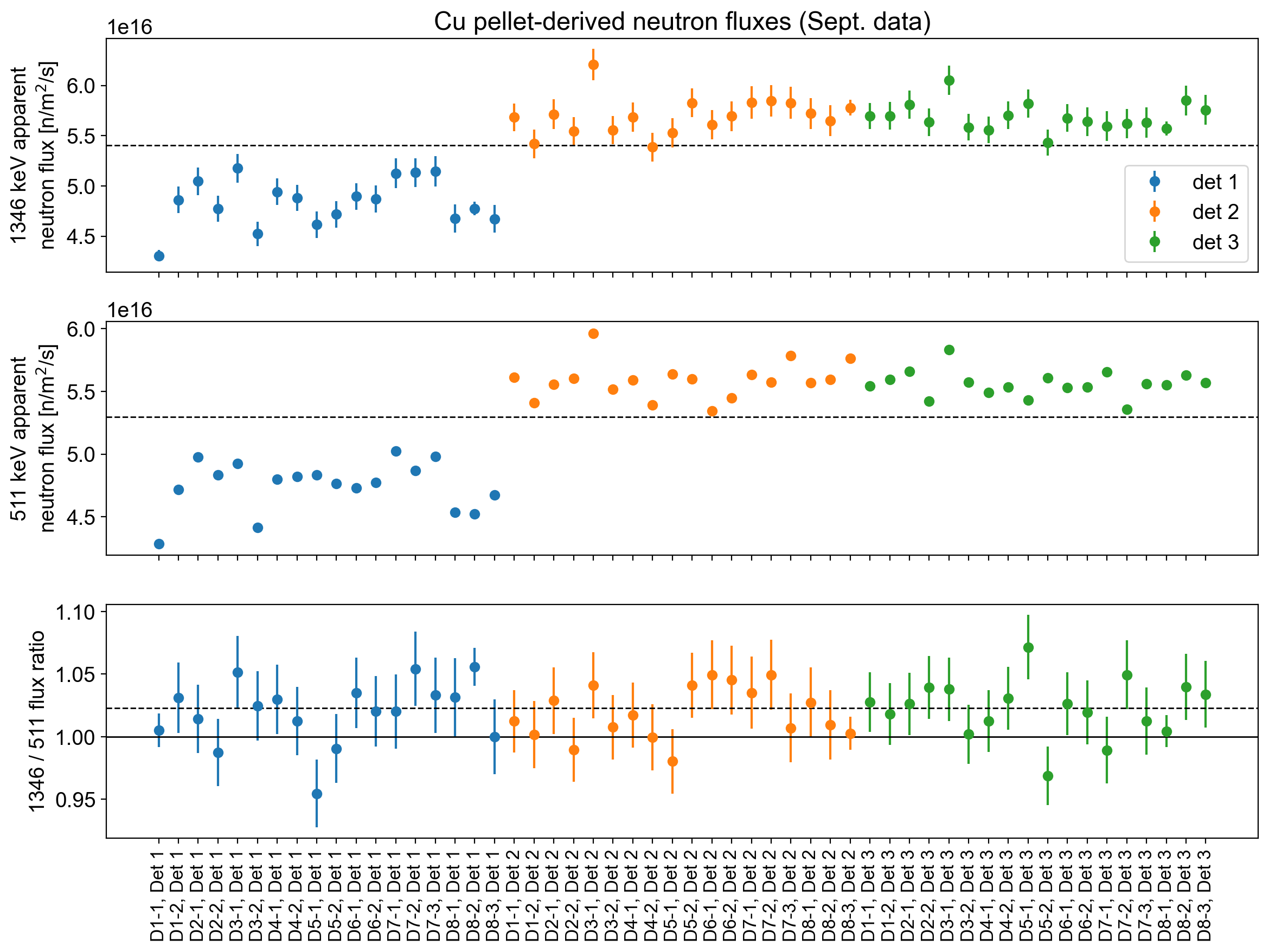}
    \caption{
        Apparent fluxes derived for each of the $18$ individual September pellets, using three different HPGe detectors, using either the $1346$~keV peak (top) or the $511$~keV peak of Cu-64 (middle).
        Dashed lines show the averages of the mean apparent fluxes for each energy, $5.407 \times 10^{16}$~neutrons/m$^2$/s and $5.299 \times 10^{16}$~neutrons/m$^2$/s for $1346$ and $511$~keV, respectively.
        The $1346/511$ ratio is also shown (bottom), and is systematically slightly higher than $1$.
    }
    \label{fig:apparent_fluxes}
\end{figure}

Mean apparent fluxes $\phi$ determined via Eq.~\ref{eq:flux_apparent} for each September pellet are shown in Fig.~\ref{fig:apparent_fluxes}.
Averaging over pellets, the mean and standard deviation of fluxes $\langle \phi \rangle$ computed from the $1346$~keV line are $(5.407 \pm 0.441) \times 10^{16}$~neutrons/m$^2$/s, or $(5.299 \pm 0.418) \times 10^{16}$~neutrons/m$^2$/s from the $511$~keV line.
We note that the standard deviations shown are the variations across pellets, and not the counting statistics uncertainty.
The reason for lower fluxes computed using one of the three HPGe detectors (detector~1) is unknown, but included as a systematic uncertainty in Section~\ref{sec:discussion}.
The resulting average activity values for the start (0800 PDT) of each flight day, as computed using the $511$~keV line, are shown in Table~\ref{tab:source_info}.
Average activity values computed with the $1346$~keV line are consistent with those from the $511$~keV line to within ${\sim}2\%$.

\renewcommand{\arraystretch}{1.1}
\setlength{\tabcolsep}{3pt}
\begin{table}[!htbp]
    \centering
    \caption{Summary of copper pellet activities}
    \begin{tabular}{c|c|c|c|c}
        date in & src configs & pellet mass & pellet mass & pellet activity$^*$ \\
        Aug.~'21 & deployed & (avg) [g] & (std dev) [\%] & (avg) [mCi] \\\hline
        9 & \begin{tabular}{c} square\\ \Lshape{} \end{tabular} & $0.2173$ & $2.43$ & $8.466$ \\\hline
        11 & \begin{tabular}{c} $8$ m separation\\ $12$ m separation\\plume \end{tabular}& $0.1746$ & $2.00$ & $6.721$ \\\hline
        13 & \begin{tabular}{c} hot/coldspot\\linear gradient\\hot line \end{tabular}& $0.1752$ & $3.37$ & $6.810$ \\
    \end{tabular}\\
    \begin{flushleft}
    $^*$~at 0800 PDT of each experiment day.\\
    \end{flushleft}
    \label{tab:source_info}
\end{table}

\section{Source deployment}

The WSU measurement campaign was conducted at an outdoor rugby field (GPS $46.7346^\circ, -117.1474^\circ$) near the WSU Nuclear Science Center from August~8 to 13, 2021, with source measurements on August~9, 11, and 13.
Source positions on the field were set out with marking flags prior to each source measurement day.
The bottom left corner $(x, y) = (65, 10)$~m common to all sources was measured via tape measure from the southwest corner of the field.
The bottom right corner was marked out in a similar fashion, after which three tape measures were used to triangulate further source boundaries.
The remainder of the source flags were then placed at $4$~m intervals between boundaries, using a tape measure to check distance and a taut string to check linearity.
This process was used to first place flags for the $10 \times 10$ source, which contains many of the source positions of the remaining seven sources, and then repeated as necessary to extend the grid for the \Lshape{}, hot line, and rectangle separation sources.
A similar process was used to place the $2$~m-spaced flags for the plume source.
Based on cross-checks of the diagonal distances, we estimate the flags were placed with a maximum error of ${\sim}4$~cm over ${\sim}40$~m, with most flags accurate to less than half that error.
Retroreflective vinyl and traffic cones were also placed at the corners of each source distribution to help localize the source extent in the measured LiDAR point clouds.
The LiDAR units measure the strength of their $905$~nm laser returns, thus including these strongly-returning retroreflective materials provides more spatial context for the measurements.

\begin{figure}[!htbp]
    \centering
    \includegraphics[width=1.0\columnwidth]{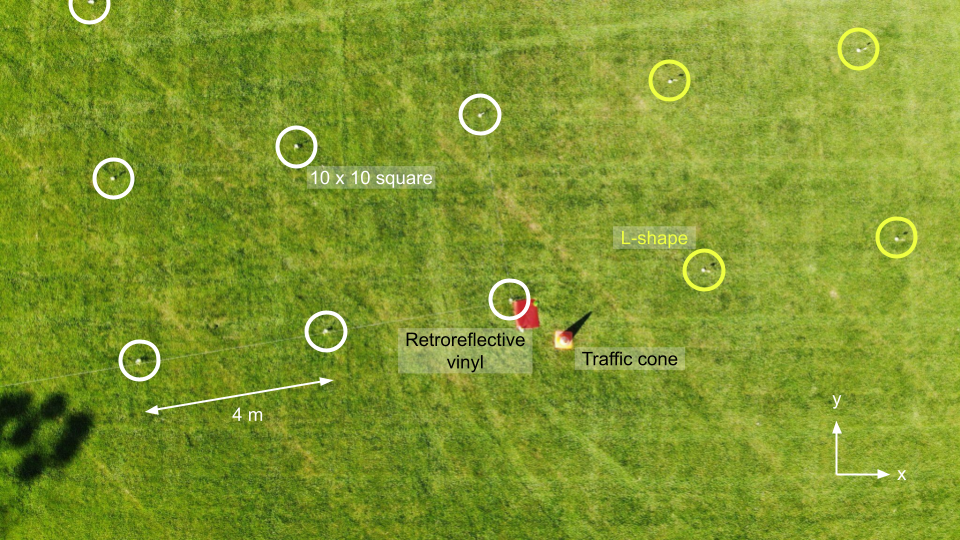}
    \caption{
        Annotated aerial photograph during the deployment of the \Lshape{} source.
        The white circles mark the source positions at the bottom right corner of the $10 \times 10$ square source that are also part of the \Lshape{}, while the yellow circles mark the positions that are only part of the \Lshape{}.
        Tennis balls containing sources are visible within the circles upon close inspection.
        The retroreflective red vinyl and orange traffic cone are also annotated.
        The dark blotches in the lower left are shadows of nearby floodlight structures.
    }
    \label{fig:aerial}
\end{figure}

On source measurement days, the sources were deployed by a team of ${\sim}10$ personnel equipped with long grabber tools to keep the radiation dose to any one person as low as reasonably achievable.
Source tennis balls were placed flush with the stem of the flag when possible, but their alignment with the grid was generally not consistent.
As a result, each true source position lies at approximately one tennis ball radius (measured to be ${\sim}3.3$~cm, consistent with the $3.27$--$3.43$~cm specified by the International Tennis Federation~\cite{itf_rules}) from the flag stem.
The tennis balls are, however, visible in aerial photogrammetry---see Fig.~\ref{fig:aerial}---allowing for small position corrections if necessary.
Given the $4$~m source separation and the goal of reconstructing distributed sources, such uncertainty is expected to be negligible.

\section{Results}\label{sec:results}
\begin{figure}[!htbp]
    \centering
    \includegraphics[width=1.0\columnwidth]{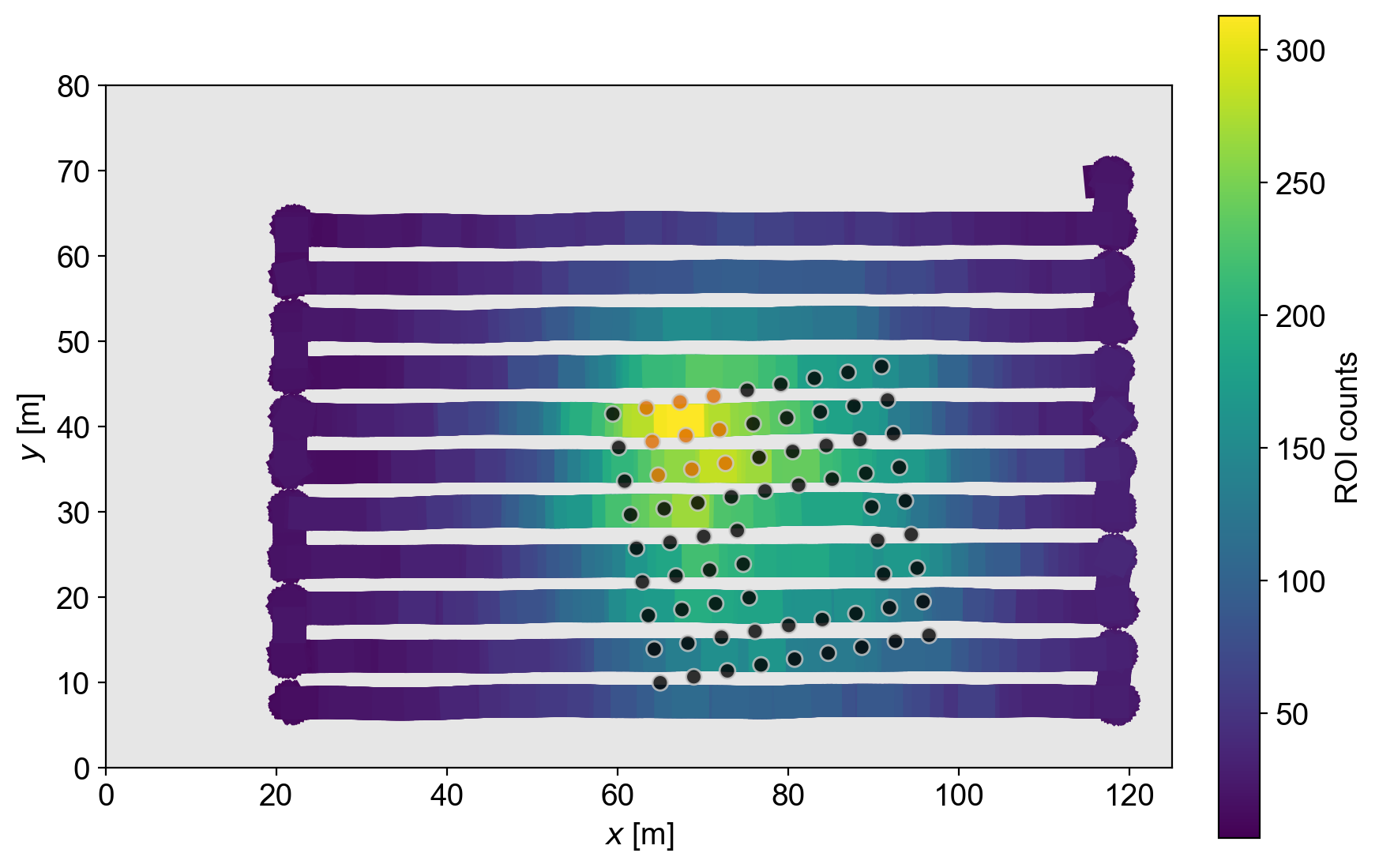}\\
    \includegraphics[width=1.0\columnwidth]{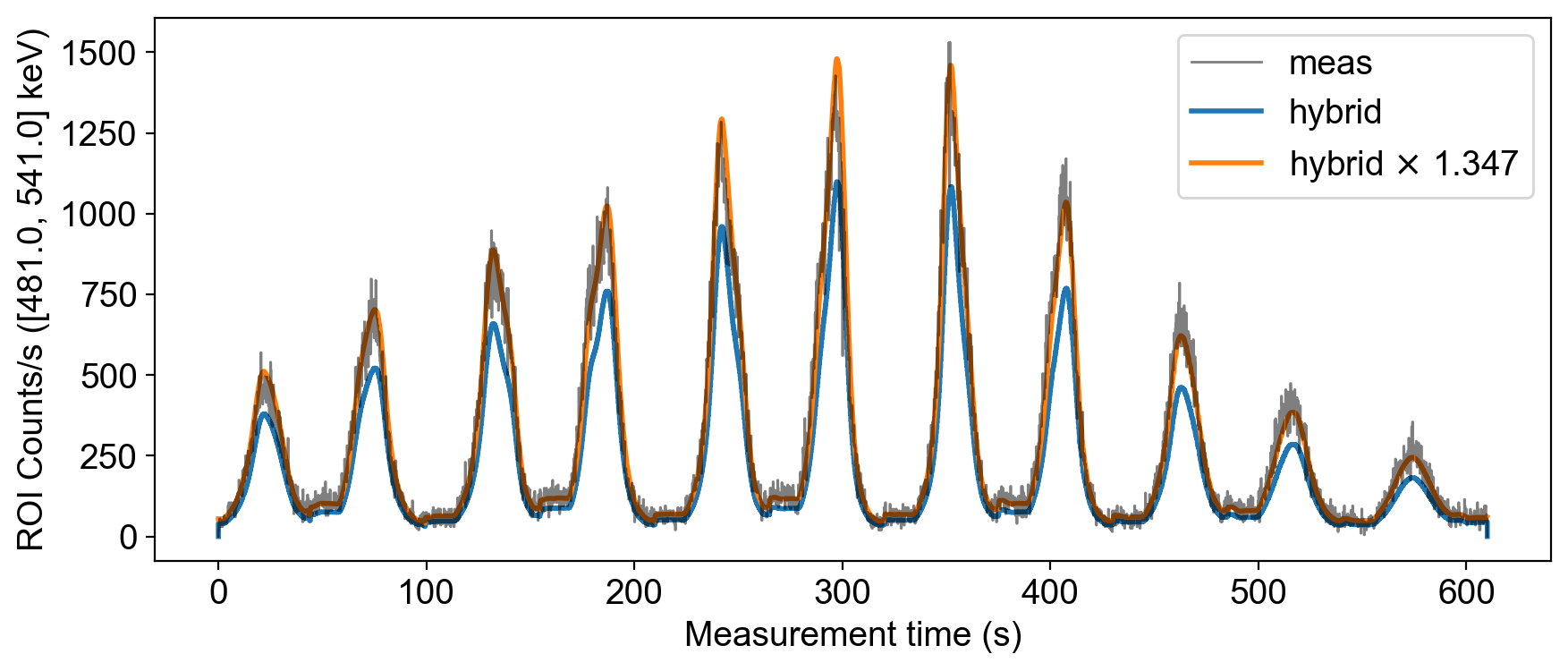}\\
    \includegraphics[width=1.0\columnwidth]{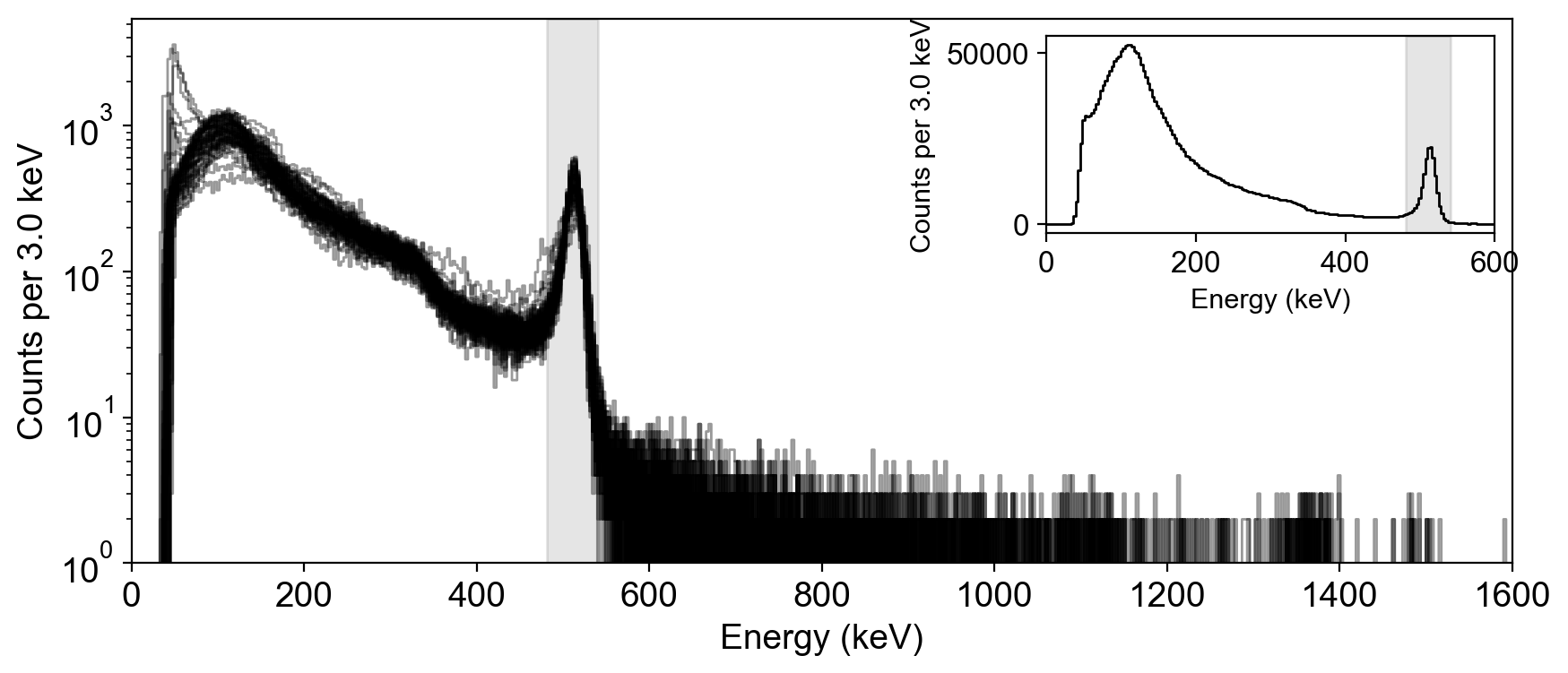}
    \caption{
        Top: top-down view of a MiniPRISM measurement $8$~m AGL over the hot/coldspot source.
        The thick curve shows the MiniPRISM detector trajectory (colorized by the measured ROI counts at each pose), over the (synthetic) source (black and orange points, as also shown in Fig.~\ref{fig:all_patterns}).
        Center: Count rate vs.\ time, summed over detector elements.
        The black histogram shows the measured counts in the $511$~keV photopeak ROI, while the blue histogram shows the ``hybrid'' expected counts computed by forward projecting the synthetic source onto the measured trajectory.
        The orange curve shows the hybrid curve scaled up by the constant scale factor of $1.347$, which was found to most closely match the data.
        Bottom: energy spectra, separated by detector elements and shown on a log scale.
        The shaded region shows the $511$~keV photopeak ROI.
        Bottom inset: energy spectrum, summed over detector elements and shown on a linear scale.
    }
    \label{fig:hotcoldspot_miniprism}
\end{figure}

\begin{figure}[!htbp]
    \centering
    \includegraphics[width=1.0\columnwidth]{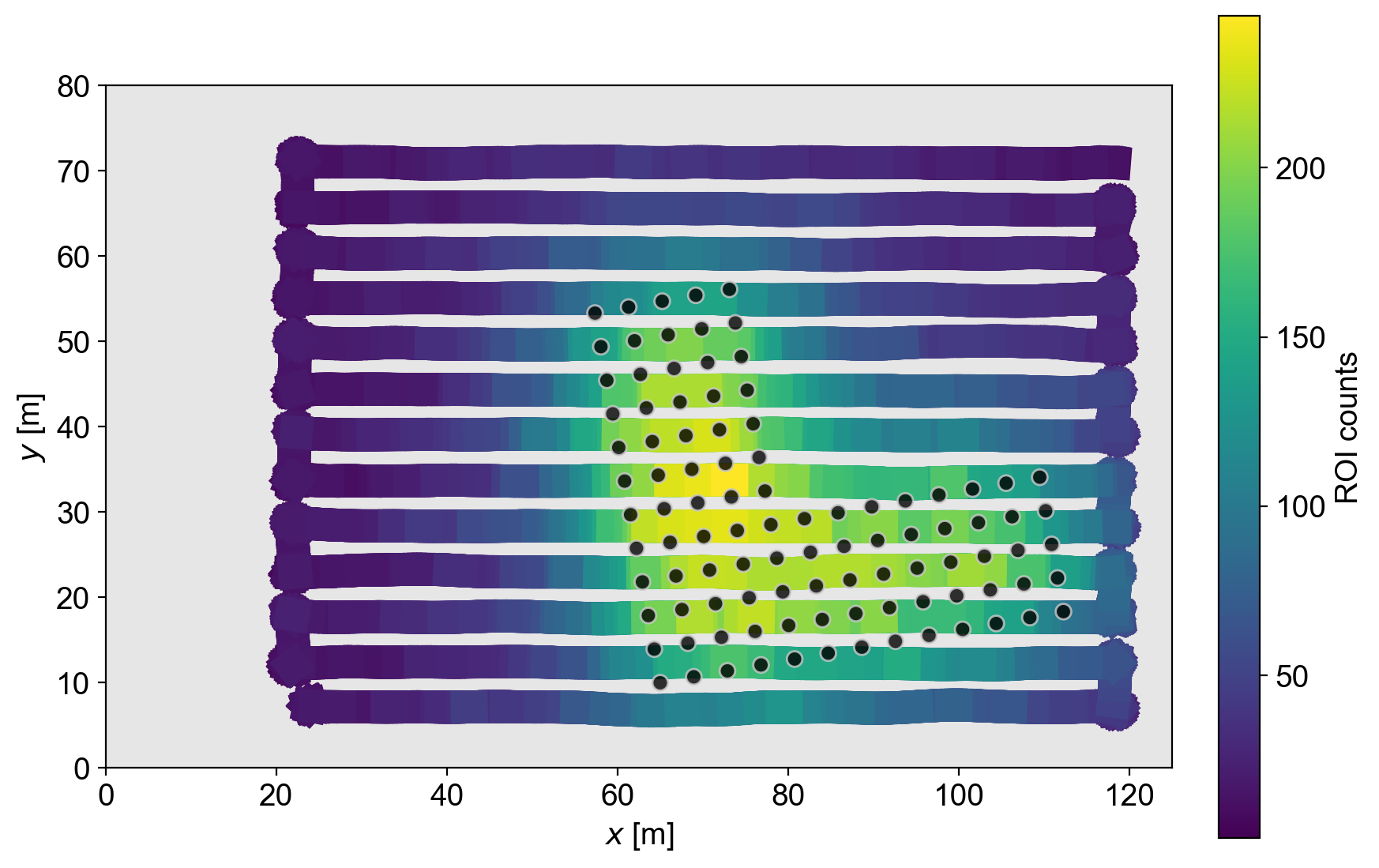}\\
    \includegraphics[width=1.0\columnwidth]{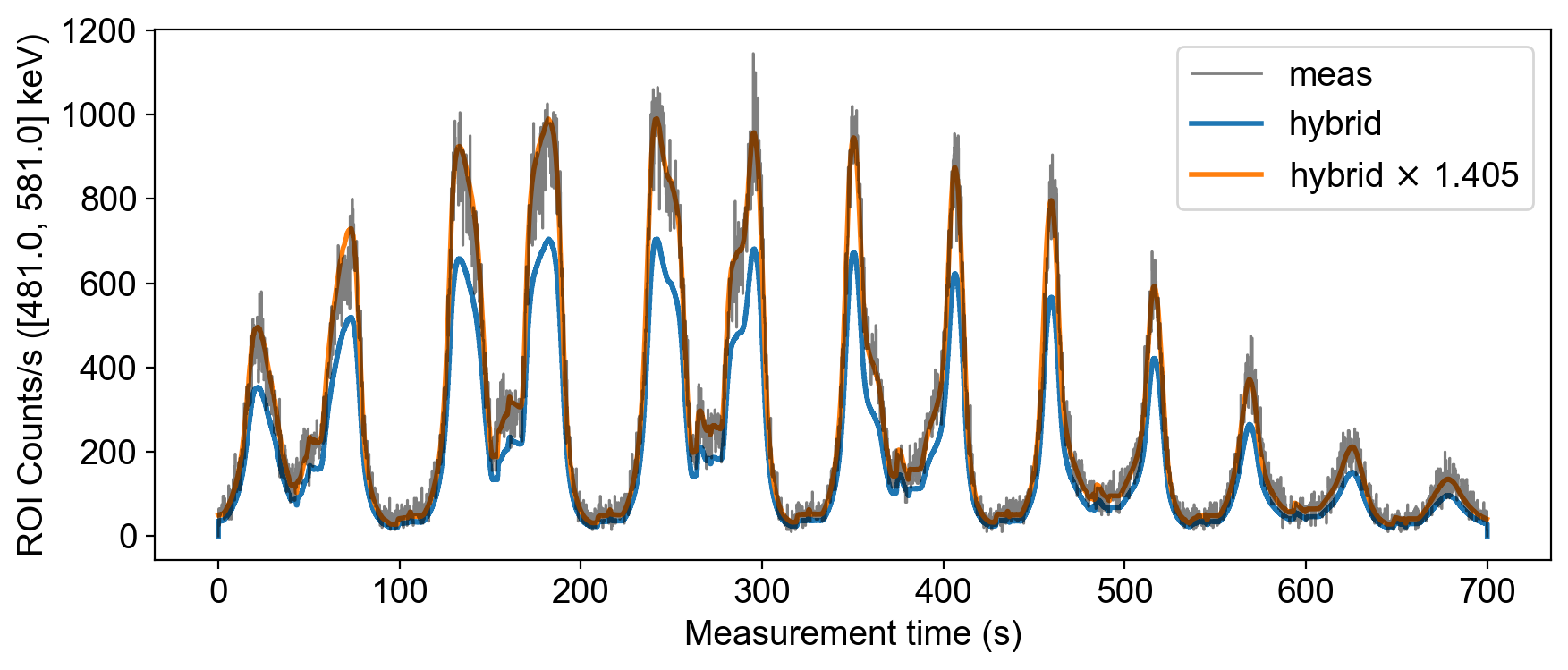}\\
    \includegraphics[width=1.0\columnwidth]{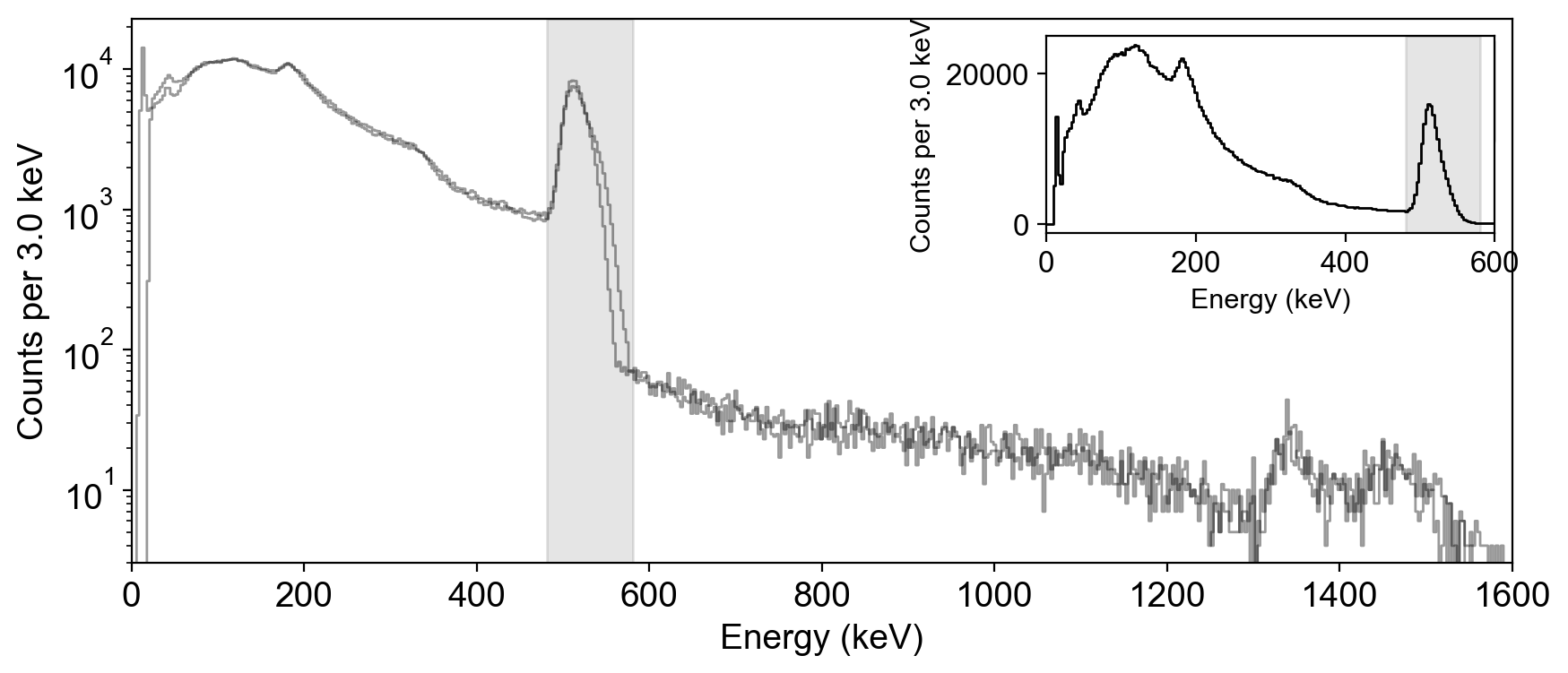}
    \caption{
        As Fig.~\ref{fig:hotcoldspot_miniprism}, but for an NG-LAMP measurement of the \Lshape{} configuration at a height of 6~m AGL.
    }
    \label{fig:Lshape_nglamp}
\end{figure}

Fig.~\ref{fig:hotcoldspot_miniprism} shows the results of an $8$~m (AGL) MiniPRISM raster over the hot/coldspot source shown in Fig.~\ref{fig:all_patterns}.
Counts in the $511$~keV photopeak ROI are plotted vs.\ position and vs.\ time with a time binning of $\Delta t = 0.2$~s; even without performing a MAP-EM reconstruction, this method of visualization produces the overall rotated square shape and hotspot.
Without the reconstruction, however, the coldspot is not easily discernable from the rest of the square.
Moreover, this visualization is not quantitative in terms of source activity.
Strong modulations in ROI counts are visible in the count rate vs.\ time plot, indicating good contrast between regions near and far from the source distribution.
Expected count rates shown in Fig.~\ref{fig:hotcoldspot_miniprism} are computed using the methods of Section~\ref{sec:forward_proj_det}, additionally accounting for air attenuation and for decay corrections from the initial activities in Table~\ref{tab:source_info} as per Ref.~\cite[Appendix~C]{knoll2010radiation}.
The measured and expected count rates across the measurement agree to $35\%$.
Possible reasons for the residual discrepancy are discussed further in Section~\ref{sec:discussion}.\footnote{The quantitative agreement between measured and expected count rates across the measurement is determined assuming there is a constant scale factor between the two, and computing the optimum scale factor that minimizes the $\chi^2$ between the scaled expected values and the data.
This method generally agrees with the ratio of the sum of counts to ${\sim}1\%$.}
In the measured spectra (shown in Figures~\ref{fig:hotcoldspot_miniprism} and~\ref{fig:Lshape_nglamp}), the $511$~keV photopeak has good contrast vs.\ background; small clusters of counts corresponding to the $1346$~keV line from the Cu-64 sources and the $1461$~keV line from the K-40 background are also visible.

Similarly, Fig.~\ref{fig:Lshape_nglamp} shows the results of a $6$~m (AGL) NG-LAMP raster over the \Lshape{} source shown in Fig.~\ref{fig:all_patterns}.
Again, the overall \Lshape{} is visible, but in this counts vs.\ position visualization, the shape is blurred over a much larger region than the true source distribution.
Due to the larger extent and lower maximum concentration of the source, the contrast in the count rate vs.\ time plot (again, $\Delta t = 0.2$~s) is not as strong as in Fig.~\ref{fig:hotcoldspot_miniprism}.
The agreement between measured and expected count rates is $41\%$, similar to the $35\%$ of Fig.~\ref{fig:hotcoldspot_miniprism}.
In the detector spectra, the $511$~keV and $1346$ Cu-64 peaks are again prominent above background, and the K-40 $1461$~keV peak and La-138 $1436$~keV self-activity peak in the CLLBC are clearly visible though not separable.

We note that several post-processing steps were performed on the data.
Although radiation measurements were collected during the entire UAS flight, the takeoff and landing segments were cut from Figs.~\ref{fig:hotcoldspot_miniprism} and \ref{fig:Lshape_nglamp} in order to focus on the constant-altitude raster pattern measurements.
Moreover, the energy spectra of Figs.~\ref{fig:hotcoldspot_miniprism} and \ref{fig:Lshape_nglamp} originally exhibited gain shifts that varied across individual detector elements, broadening the summed photopeak and shifting it away from $511$~keV.
To compensate, we find the per-detector linear gain shift necessary to return the photopeak to 511 keV, and apply that to the detector's energy data.
To avoid aliasing the energy data in the process, we also apply a zero-mean Gaussian blur of $1$~keV standard deviation to the listmode energy data, which we note is much less than the expected photopeak standard deviations of ${\mathcal{O}}(10$~keV$)$.
Similarly, for NG-LAMP, a small time noise was applied to plots of the ROI counts vs.\ measurement time to avoid time bin aliasing.
Some NG-LAMP and MiniPRISM detector elements were suffering from large amounts of electronic noise or were not reading out data, respectively, during several of the UAS experiments; in these cases, data from the problematic detector elements and the corresponding contributions to computed sensitivity are excluded from the analysis.

Coordinate transforms between the measured data and the idealized field coordinate system (see Section~\ref{sec:example_calcs}) used throughout this work were determined in two steps---RTK $\leftrightarrow$ SLAM then SLAM $\leftrightarrow$ field---since there are no clear features in the RTK trajectories that reliably map onto the field coordinate system.
First, we aligned a UAS trajectory reconstructed by LiDAR SLAM to the same trajectory as measured by the RTK to obtain RTK $\leftrightarrow$ LiDAR transforms.
The optimal alignment was obtained by minimizing the sum of squared distances between the two sets of trajectory coordinates (interpolated to the same timestamps) over four parameters, the $xyz$ translation and yaw.
The LiDAR $\leftrightarrow$ field transforms were then estimated by picking points in the LiDAR point clouds at the western corners of the field boundary and aligning their vector with the $+y$ axis of the field coordinate system.


\section{Discussion}\label{sec:discussion}

The results of Section~\ref{sec:results}---in particular, Figs.~\ref{fig:hotcoldspot_miniprism} and~\ref{fig:Lshape_nglamp}---show $35$--$40\%$ agreement between the measured and expected 511~keV photopeak count rate.
Agreement to this level is in fact seen across the entire set of measurements (in which RTK position data is available), and is relatively constant across measurement days, source configurations, and detector systems.
In particular, the average levels of agreement for the NG-LAMP, MiniPRISM, and overall datasets are $41\%$, $32\%$, and $38\%$, respectively.
For completeness, though, we note that we have analyzed and either corrected for or ruled out several possible sources of error, and made estimates of our dominant systematic uncertainties:
\begin{enumerate}
    \item \textbf{attenuation in air:}
    although most flight altitudes were ${\lesssim}10$~m, the distributed nature of the source and wide raster patterns mean that source-to-detector distances are often on the order of the $511$~keV mean free path in air, $96.2$~m for dry air near sea level~\cite{hubbell1995tables}.
    Air attenuation therefore causes non-negligible reductions in the expected photopeak counts.
    To model air attenuation, we use the NIST XCOM mass attenuation coefficient tables for dry sea-level air~\cite{nist_xcom}, but choose an air density that reflects local weather conditions in Pullman, WA on the measurement days.
    We use an air temperature of $28\,^\circ$C, pressure of $928$~hPa, and relative humidity of $30\%$ as representative values across and within all three days~\cite{pullman_weather}.
    We then use the simplified air density formula of Ref.~\cite{air_density_calc} to arrive at an air density of $\rho = 1.073$~g/cm$^3$.
    Our forward projections then correct Eq.~\ref{eq:lambda_arr} for air attenuation between each point-source position $\pvec{r}_k$ and detector position $\vec{r}_i$.
    For the measurements shown in Figs.~\ref{fig:hotcoldspot_miniprism} and \ref{fig:Lshape_nglamp}, the overall magnitude of the correction is an approximately $17\%$ reduction in the sum of the hybrid photopeak counts.
    As we have used a single air density value, changes in pressure and temperature throughout the week of measurements are expected to induce a ${\sim}3\%$ uncertainty in forward-projected counts.

    \item \textbf{attenuation in the source holder:}
    the activities of Table~\ref{tab:source_info} were computed for bare copper pellets without the source holders (epoxy encasing and tennis balls).
    To correct for attenuation losses in the source holders, we simulate monoenergetic photons emitted isotropically and uniformly throughout a $0.19$~g copper pellet of diameter and height $3$~mm.
    The copper volume is centered in an epoxy cylinder of diameter $1.5$~cm and length $4$~cm, surrounded by air, and placed inside a tennis ball volume comprising a $3.5$~mm rubber shell and a $2.5$~mm felt shell.
    Material compositions were generally taken from Ref.~\cite{detwiler2021compendium}, though high-density ($1.5$~g/cm$^3$) rubber~\cite{sissler2010viscoelastic} was used to ensure the modelled tennis ball mass was consistent with its measured value of $57$~g.
    Since both the on-field and HPGe assay measurements are subject to attenuation by the copper pellet itself, we define the transmission fraction as the ratio of full-energy photons escaping the tennis ball to those escaping the copper pellet, thereby quantifying losses from the epoxy, internal air, and tennis ball volumes only.
    Moreover, since the orientations of the epoxy cylinders were randomized during the field measurements, the transmission fraction is taken as an average over all emission directions.
    We find that the attenuation losses are primarily driven by the epoxy cylinder, and that the average transmission fractions are $0.861$ for $511$~keV photons or $0.910$ for $1346$~keV.
    We then multiply the pellet activities in Table~\ref{tab:source_info} by $0.861$ to obtain apparent source activities for the forward projections in Figs.~\ref{fig:hotcoldspot_miniprism} and \ref{fig:Lshape_nglamp}.
    We note however that while we have assumed the epoxy orientations were isotropic, the vials may have tended to settle closer to a horizontal orientation, thereby biasing the detected emission distribution away from the poles and reducing the epoxy attenuation while the detector was overhead.
    In the extreme case of all emissions occurring in the radial direction, the transmission fraction would increase to about $0.89$ for $511$~keV photons.
    The increase in transmission due to the epoxy orientation bias is therefore at most $3\%$, and likely much less.
    
    \item \textbf{511~keV region of interest:} care must be taken when defining the $511$~keV region of interest (ROI) in the HPGe ground truth activity assay.
    We found that the default $511$~keV ROI provided by the Genie™ 2000 spectroscopy software~\cite{genie2kanalysis} relied on a peak width calibration based on nuclear decay lines.
    However, the $511$~keV line is subject to additional Doppler broadening~\cite[p.~441]{knoll2010radiation}, rendering its width substantially larger than nearby decay lines.
    Because the same Genie™ 2000 analysis pipeline was used to empirically plan irradiation times and to assay the pellets used, this too-narrow ROI led to both a ${\sim}30\%$ overproduction of Cu-64 and a corresponding initial ${\sim}30\%$ underestimate of the Cu-64 ground truth activity.
    In our final analyses, this ROI is broadened to include the entire $511$~keV peak.
    
    \item \textbf{in-scatter:} although we have corrected our forward projections (Eq.~\ref{eq:lambda_arr}) for scattering losses between the sources and the detector, we have so far only considered scattering as a loss mechanism.
    However, due to the often-large source-to-detector standoffs and the non-zero width of the energy ROI used to define the measured $511$~keV signal, there is a non-negligible solid angle in which small-angle in-scatter (``buildup'', typically from air, but potentially also from the ground) can occur, increasing the ROI signal compared to the prediction of Eq.~\ref{eq:lambda_arr}.
    This effect manifests as the apparent step function underneath the $511$~keV peak (most notably in Fig.~\ref{fig:Lshape_nglamp}) since Compton scattering always reduces the photon energy.
    To compensate for this in-scatter without running computationally intensive scattering simulations,\footnote{The higher-fidelity and only moderately more computationally intensive air scattering model outlined in Ref.~\cite{salathe2021determining} could be useful here, but would require substantial work to implement on the GPU.} we derive an approximate buildup correction from the global $511$~keV peak (i.e., summed over the entire run) from the NG-LAMP run in Fig.~\ref{fig:Lshape_nglamp}.
    We fit the peak with an exponentially-modified Gaussian peak shape plus two different exponential backgrounds on either side of the centroid that are blurred together by convolution with the peak's resolution.
    We find that the area under this ``source-induced background'' component comprises approximately $9.2\%$ of the ROI area.
    Although this exact value will depend on the source geometry, trajectory, and perhaps even the detector, we take it as a representative in-scatter fraction for our experiments.
    In the comparisons of Figs.~\ref{fig:hotcoldspot_miniprism} and \ref{fig:Lshape_nglamp}, we therefore downsample the measured listmode data, randomly dropping $9.2\%$ of events.

    \item \textbf{altitude errors:}
    when dealing with a single point source, small source-to-detector distance errors have a quadratic effect on the expected counts $\bvec{\lambda}$.
    However, as shown in Eq.~\ref{eq:lambda_plane}, for distributed sources the dependence is generally logarithmic, and thus it would take altitude errors of several meters to explain the discrepancy between our measurements and expectation based on the analysis of the activity assays.
    Moreover, adjusting the detector altitude scales counts non-linearly with respect to source proximity, generally improving the agreement in only some parts of the count rate comparisons while worsening it in others.
    There is a small altitude uncertainty of around ${\pm}25$~cm due to the non-zero slope of the field and the accuracy of the spatial transformations, corresponding to a roughly ${\pm}5\%$ change in expected counts.

    \item \textbf{detector pitch:} the measurements with RTK trajectory data do not contain UAS pitch or roll info, but the detector angular response varies with system orientation.
    We find that adding a constant $10^\circ$ pitch, representative of real pitches observed on the field, would change the summed forward projected counts by ${<}1\%$.
    In the extreme case of a constant $45^\circ$ pitch, the change would be ${\sim}2\%$.
    For simplicity, therefore, no pitch was imputed to the RTK trajectories in the analyses presented.

    \item \textbf{detector response validation:} we showed good agreement between data and our response models for the MiniPRISM detector in Ref.~\cite{hellfeld2021free}.
    We performed an additional validation study with NG-LAMP and Na-22 check sources (which also emit at $511$~keV), and found agreement between expected and measured effective areas to within ${\sim}10\%$.
    Moreover, the average difference of $8\%$ between the NG-LAMP and MiniPRISM datasets in this work is consistent with the fidelity to which we expect to know our detector responses.
    Finally, we note that our detector response models do not use $511$~keV response simulations directly, but rather interpolate (in log-log space) between two nearby energies ($356$ and $662$~keV for NG-LAMP, and $500$ and $600$~keV for MiniPRISM).
    Because the response efficiencies decrease roughly exponentially above ${\sim}200$~keV, performing the energy interpolation in linear space instead of log-log space would lead to a $+15\%$ interpolation error with NG-LAMP (but only $+1\%$ with MiniPRISM).

    \item \textbf{forward projection code:}
    we compared the GPU-based {\tt mfdf} forward projection code (Eq.~\ref{eq:lambda_arr}) against a separate simpler CPU-based Python implementation and found results consistent to ${\sim}1\%$.
    Moreover, we modelled with {\tt mfdf} one of the distributed Na-22 $511$~keV sources in Ref.~\cite{hellfeld2021free}, and replicated the measured counts to within experimental error.

    \item \textbf{decay correction to mean activity:}
    given that the ${\sim}15$~minute measurement times are not entirely insignificant (${\sim}1\%$) compared to the Cu-64 half-life ($t_{1/2}=12.7$~hours), we decay-correct the source activity in each measurement to the mean activity over the measurement time (following Ref.~\cite[Appendix~C]{knoll2010radiation}) rather than simply the activity at, e.g., the start or midpoint time.

    \item \textbf{indirect production of 511~keV photons:}

    \begin{enumerate}
        \item pair production and scatter from the $1346$~keV line in the detector and surrounding materials: as an extreme upper limit, even if every $1346$~keV photon incident on the detector led to a $511$~keV detection, the relative contribution to the $511$~keV ROI would be limited to about $0.25\%$ based on photon yield~\cite{singh2007nuclear} and efficiency ratios.
        \item pair production and scatter from the $1346$~keV line in the ground: we find via Monte Carlo simulation that for an isotropically-emitting plane source, each $1346$~keV photon results in only ${\sim}6\times 10^{-3}$ photons leaving the ground in a $511 \pm 30$~keV energy window.
        \item pair production and scatter from the $1346$~keV line in the source holder: based on our simulations of attenuation in the epoxy and tennis ball source holders, fewer than $10^{-3}$ annihilation photons are emitted from the tennis ball for every $1346$~keV photon generated in the copper pellet.
    \end{enumerate}
\end{enumerate}
Altogether, our dominant known sources of systematic uncertainty are as follows:
\begin{itemize}
    \item detector response: ${\sim}10\%$.
    \item Cu-64 source activities and associated mean apparent flux calculations: ${\sim}10\%$ from the observed spread in activities across the three HPGe detectors.
    \item in-scatter: ${\sim}10\%$ from the approximate nature of the global in-scatter correction.
    \item altitude uncertainties: ${\sim}5\%$ from potential detector altitude errors.
    \item air density uncertainties: ${\sim}3\%$ from using a single representative air density for all measurements.
\end{itemize}
Potential unknown sources of systematic uncertainty, such as the epistemic uncertainty in the source activity re-determination of Section~\ref{sec:assay}, are not quantified.
Adding these known systematics linearly we have a total systematic error of ${\sim}38\%$, consistent with the observed $35\%$ and $41\%$ excess counts observed in Figs.~\ref{fig:hotcoldspot_miniprism} and \ref{fig:Lshape_nglamp}, and with the average excesses of $41\%$, $32\%$, and $38\%$ observed in the NG-LAMP, MiniPRISM, and overall datasets.
To improve these systematics in potential future measurement campaign, we recommend using a true nuclear decay line (e.g., the $662$~keV line of Cs-137 rather than $511$~keV) to avoid spectroscopy errors; ensuring the same set of sources that is used during the measurement is used for the source activity assay; and improving air models for both the air in-scatter and weather-dependent attenuation corrections.
Finally, we note that while this $35$--$40\%$ discrepancy between measurements and modeling exists, since it appears to be fairly consistent across detectors and both across and within measurements, we can still quantitatively map distributed radiological sources, though some absolute activity scale correction or calibration may be required.

In general, our point-source array technique serves as a useful proof-of-concept for future measurement campaigns where well-controlled distributed sources would be desirable.
We expect these techniques to be useful up to source dimensions of around $100$~m, beyond which it may be preferable to develop a more scalable method at the expense of some spatial accuracy.

Having a team of ${\sim}10$ personnel to deploy sources at the pre-set flags was instrumental in deploying, reconfiguring, and removing source distributions on the order of ten minutes, keeping dose to personnel relatively low in the process.
The Berkeley personnel, who were often present on the field to place sources and pilot or spot the UAS, received an average (standard deviation) full-body dose equivalent of $220(17)$~{\textmu}Sv over the entire measurement campaign.
We expect this dose could be further reduced in a truly open-field measurement with no surrounding fences or light poles, as the UAS team could safely increase their standoff to the measurement area.

We also note that for operational simplicity, we have only demonstrated point-source array designs and measurements on a flat 2D ground plane, and generally with fixed-altitude raster patterns above the source.
Generalizations to arrays over hilly surfaces or full 3D environments are possible, but will require a map of the environment to be modeled or measured (e.g., via LiDAR SLAM).
In fact when such a map is available, it is possible to account for attenuation in the scene~\cite{bandstra2021improved} that has made activity reconstruction difficult in similar measurement scenarios~\cite{murtha2021tomographic}.

Finally, we emphasize again that analyzing detected counts vs.\ $xy$ position as in Figs.~\ref{fig:hotcoldspot_miniprism} and~\ref{fig:Lshape_nglamp} does not provide quantitative estimates of the source total activity or distribution.
To answer such questions, we require quantitative reconstruction methods.
In an upcoming work, we will use regularized ML-EM (MAP-EM) to quantitatively reconstruct the source distributions and compare to known ground truth; such work could also involve other methods under study such as particle filters and genetic algorithms.

\section{Conclusion}
We have demonstrated a method for emulating distributed gamma ray sources with arrays of sealed point sources, and performed aerial measurements of Cu-64 source arrays of total activity up to ${\sim}700$~mCi.
We found that the sealed array sources were easily reconfigurable, enabling multiple source configuration measurements per day, and we obtained quantitative agreement between modelled and measured count rates at the level of ${\lesssim}40\%$, consistent with our expected systematic uncertainties.
This agreement could be improved in potential future measurement campaigns by enhancing the precision of the ground truth activity assay and by better understanding second-order effects such as air in-scatter.
These measurements will form the basis of upcoming studies applying quantitative reconstruction techniques to accurately determine the shape and magnitude of distributed radiological sources.

\section*{Acknowledgements}
\small
This material is based upon work supported by the Defense Threat Reduction Agency under HDTRA 13081-36239.
This support does not constitute an express or implied endorsement on the part of the United States Government.
Distribution~A: approved for public release, distribution is unlimited.

This document was prepared as an account of work sponsored by the United States Government. While this document is believed to contain correct information, neither the United States Government nor any agency thereof, nor the Regents of the University of California, nor any of their employees, makes any warranty, express or implied, or assumes any legal responsibility for the accuracy, completeness, or usefulness of any information, apparatus, product, or process disclosed, or represents that its use would not infringe privately owned rights. Reference herein to any specific commercial product, process, or service by its trade name, trademark, manufacturer, or otherwise, does not necessarily constitute or imply its endorsement, recommendation, or favoring by the United States Government or any agency thereof, or the Regents of the University of California. The views and opinions of authors expressed herein do not necessarily state or reflect those of the United States Government or any agency thereof or the Regents of the University of California.

This manuscript has been authored by an author at Lawrence Berkeley National Laboratory under Contract No.~DE-AC02-05CH11231 with the U.S.~Department of Energy. The U.S.~Government retains, and the publisher, by accepting the article for publication, acknowledges, that the U.S.~Government retains a non-exclusive, paid-up, irrevocable, world-wide license to publish or reproduce the published form of this manuscript, or allow others to do so, for U.S.~Government purposes.

This research used the Lawrencium computational cluster resource provided by the IT Division at the Lawrence Berkeley National Laboratory (Supported by the Director, Office of Science, Office of Basic Energy Sciences, of the U.S.\ Department of Energy under Contract No.\ DE-AC02-05CH11231).

The authors gratefully acknowledge Erika Suzuki and Gamma Reality, Inc.~(GRI) for providing the aerial photograph in Fig.~\ref{fig:aerial}, and Ivan Cho for piloting the UAS during several measurements.

The authors also gratefully acknowledge the remote assistance provided during both the measurement campaign and the analysis by Joseph Curtis, Joshua Cates, Ryan Pavlovsky, and especially Marco Salathe, all of Lawrence Berkeley National Laboratory, as well as support from the Immersive Semi-Autonomous Aerial Command System (ISAACS) group at UC Berkeley.

Finally, the authors thank the staff of the WSU Nuclear Science Center for their assistance in deploying the sources on the field.

\ifCLASSOPTIONcaptionsoff
  \newpage
\fi



%

\bibliographystyle{unsrt}
\bibliography{bib}

\end{document}